\documentclass[onecolumn,epjc3]{svjour3} 

\usepackage[space=true]{accsupp}
\usepackage{acronym}
\usepackage{amsmath}
\usepackage{amssymb}
\usepackage[american]{babel}
\usepackage{booktabs}
\usepackage{caption}
\usepackage{newfloat}
\usepackage[utf8]{inputenc}
\usepackage[left=.09\paperwidth,right=.09\paperwidth,bottom=.1\paperheight,top=.1\paperheight]{geometry}
\usepackage{graphicx}
\graphicspath{{./figures/}}
\usepackage{textcomp}

\usepackage[dvipsnames]{xcolor}
\definecolor{darkblue}{rgb}{0,0,0.9}
\usepackage{hyperref}
\hypersetup{
colorlinks,
citecolor=darkblue,
filecolor=darkblue,
linkcolor=darkblue,
urlcolor=darkblue
}

\usepackage{listings}
\usepackage[numbers,compress]{natbib}
\usepackage{orcidlink}
\usepackage{placeins}
\usepackage{subcaption}
\usepackage[]{todonotes}
\usepackage[normalem]{ulem}
\usepackage{xcolor}
\usepackage{xspace}
\makeatletter
\ifcsname c@chapter\endcsname\else
    \def\cl@chapter{\@elt{theorem}}
\fi
\makeatother
\usepackage[capitalize,nameinlink]{cleveref}

\newcommand{\noncopynumber}[1]{%
    \BeginAccSupp{method=escape,ActualText={}}%
    #1%
    \EndAccSupp{}%
}
\newcommand{\copyablespace}{\BeginAccSupp{method=hex,unicode,ActualText=00A0}\ \EndAccSupp{}}

\definecolor{maroon}{cmyk}{0, 0.87, 0.68, 0.32}
\definecolor{halfgray}{gray}{0.55}
\definecolor{ipython_frame}{RGB}{207, 207, 207}
\definecolor{ipython_bg}{RGB}{247, 247, 247}
\definecolor{ipython_red}{RGB}{186, 33, 33}
\definecolor{ipython_green}{RGB}{0, 128, 0}
\definecolor{ipython_cyan}{RGB}{64, 128, 128}
\definecolor{ipython_purple}{RGB}{170, 34, 255}

\lstdefinelanguage{iPython}{
    morekeywords={access,and,break,class,continue,def,del,elif,else,except,exec,finally,for,from,global,if,import,in,is,lambda,not,or,pass,print,raise,return,try,while},%
    morekeywords=[2]{abs,all,any,basestring,bin,bool,bytearray,callable,chr,classmethod,cmp,compile,complex,delattr,dict,dir,divmod,enumerate,eval,execfile,file,filter,float,format,frozenset,getattr,globals,hasattr,hash,help,hex,id,input,int,isinstance,issubclass,iter,len,list,locals,long,map,max,memoryview,min,next,object,oct,open,ord,pow,property,range,raw_input,reduce,reload,repr,reversed,round,set,setattr,slice,sorted,staticmethod,str,sum,super,tuple,type,unichr,unicode,vars,xrange,zip,apply,buffer,coerce,intern},%
    sensitive=true,%
    morecomment=[l]\#,%
    morestring=[b]',%
    morestring=[b]",%
    morestring=[s]{'''}{'''},
    morestring=[s]{"""}{"""},
    morestring=[s]{r'}{'},
    morestring=[s]{r"}{"},%
    morestring=[s]{r'''}{'''},%
    morestring=[s]{r"""}{"""},%
    morestring=[s]{u'}{'},
    morestring=[s]{u"}{"},%
    morestring=[s]{u'''}{'''},%
    morestring=[s]{u"""}{"""},%
    literate=
    {á}{{\'a}}1 {é}{{\'e}}1 {í}{{\'i}}1 {ó}{{\'o}}1 {ú}{{\'u}}1
    {Á}{{\'A}}1 {É}{{\'E}}1 {Í}{{\'I}}1 {Ó}{{\'O}}1 {Ú}{{\'U}}1
    {à}{{\`a}}1 {è}{{\`e}}1 {ì}{{\`i}}1 {ò}{{\`o}}1 {ù}{{\`u}}1
    {À}{{\`A}}1 {È}{{\'E}}1 {Ì}{{\`I}}1 {Ò}{{\`O}}1 {Ù}{{\`U}}1
    {ä}{{\"a}}1 {ë}{{\"e}}1 {ï}{{\"i}}1 {ö}{{\"o}}1 {ü}{{\"u}}1
    {Ä}{{\"A}}1 {Ë}{{\"E}}1 {Ï}{{\"I}}1 {Ö}{{\"O}}1 {Ü}{{\"U}}1
    {â}{{\^a}}1 {ê}{{\^e}}1 {î}{{\^i}}1 {ô}{{\^o}}1 {û}{{\^u}}1
    {Â}{{\^A}}1 {Ê}{{\^E}}1 {Î}{{\^I}}1 {Ô}{{\^O}}1 {Û}{{\^U}}1
    {œ}{{\oe}}1 {Œ}{{\OE}}1 {æ}{{\ae}}1 {Æ}{{\AE}}1 {ß}{{\ss}}1
    {ç}{{\c c}}1 {Ç}{{\c C}}1 {ø}{{\o}}1 {å}{{\r a}}1 {Å}{{\r A}}1
    {€}{{\EUR}}1 {£}{{\pounds}}1
    {^}{{{\color{ipython_purple}\^{}}}}1
    {=}{{{\color{ipython_purple}=}}}1
    {+}{{{\color{ipython_purple}+}}}1
    {*}{{{\color{ipython_purple}$^\ast$}}}1
    {/}{{{\color{ipython_purple}/}}}1
    {?}{{{\color{ipython_purple}?}}}1
    {+=}{{{+=}}}1
    {-=}{{{-=}}}1
    {*=}{{{$^\ast$=}}}1
    {/=}{{{/=}}}1
    {\ }{{\copyablespace}}1,
    identifierstyle=\color{black}\ttfamily,
    commentstyle=\color{ipython_cyan}\ttfamily,
    stringstyle=\color{ipython_red}\ttfamily,
    keepspaces=true,
    showspaces=true,
    showstringspaces=false,
    rulecolor=\color{ipython_frame},
    frame=single,
    frameround={t}{t}{t}{t},
    framexleftmargin=6mm,
    numbers=none,
    numberstyle=\tiny\color{halfgray}\noncopynumber,
    backgroundcolor=\color{ipython_bg},
    basicstyle=\scriptsize\ttfamily,
    keywordstyle=\color{ipython_green}\ttfamily,
    aboveskip=\smallskipamount,
    belowskip=\smallskipamount
}
\definecolor{maroon}{cmyk}{0, 0.87, 0.68, 0.32}
\definecolor{halfgray}{gray}{0.55}
\definecolor{bash_frame}{RGB}{207, 207, 207}
\definecolor{bash_bg}{RGB}{247, 247, 247}
\definecolor{bash_red}{RGB}{186, 33, 33}
\definecolor{bash_green}{RGB}{0, 128, 0}
\definecolor{bash_cyan}{RGB}{64, 128, 128}
\definecolor{bash_purple}{RGB}{170, 34, 255}

\lstdefinelanguage{bash}[]{sh}%
  {morekeywords={alias,bg,bind,builtin,caller,command,compgen,compopt,%
      complete,coproc,curl,declare,disown,dirs,enable,fc,fg,help,%
      history,jobs,let,local,logout,mapfile,printf,pushd,popd,%
      readarray,select,set,suspend,shopt,source,times,type,typeset,%
      ulimit,unalias,wait},%
   otherkeywords={ [, ], [[, ]], \{, \} }%
  }%

\lstdefinelanguage{bash}{
    morekeywords={awk,break,case,cat,cd,continue,do,done,echo,elif,else,%
      env,esac,eval,exec,exit,export,expr,false,fi,for,function,getopts,%
      hash,history,if,in,kill,login,newgrp,nice,nohup,ps,pwd,read,%
      readonly,return,set,sed,shift,test,then,times,trap,true,type,%
      ulimit,umask,unset,until,wait,while},%
    morecomment=[l]\#,%
    morestring=[d]",%
    alsoletter={*"'0123456789.},%
    alsoother={\{\=\}},%
    literate={{=}{{{=}}}1},%
    literate={\$\{}{{{{\bfseries{}\$\{}}}}2,%
    otherkeywords={ [, ], \{, \} },%
    identifierstyle=\color{black}\ttfamily,
    commentstyle=\color{bash_cyan}\ttfamily,
    stringstyle=\color{bash_red}\ttfamily,
    keepspaces=true,
    showspaces=false,
    showstringspaces=false,
    rulecolor=\color{bash_frame},
    frame=single,
    frameround={t}{t}{t}{t},
    framexleftmargin=6mm,
    numbers=none,
    numberstyle=\tiny\color{halfgray},
    backgroundcolor=\color{bash_bg},
    basicstyle=\scriptsize\ttfamily,
    keywordstyle=\color{bash_green}\ttfamily,
    aboveskip=\smallskipamount,
    belowskip=\smallskipamount,
}[keywords,comments,strings]%
\definecolor{maroon}{cmyk}{0, 0.87, 0.68, 0.32}
\definecolor{halfgray}{gray}{0.55}
\definecolor{yaml_frame}{RGB}{207, 207, 207}
\definecolor{yaml_bg}{RGB}{247, 247, 247}
\definecolor{yaml_red}{RGB}{186, 33, 33}
\definecolor{yaml_green}{RGB}{0, 128, 0}
\definecolor{yaml_cyan}{RGB}{64, 128, 128}
\definecolor{yaml_purple}{RGB}{170, 34, 255}

\newcommand\YAMLpunctstyle{\color{black}\mdseries}
\newcommand\YAMLpunct[1]{{\YAMLpunctstyle#1}}
\newcommand\YAMLkeystyle{\color{yaml_green}\mdseries}
\newcommand\YAMLvaluestyle{\color{blue}\mdseries}
\makeatletter
\def\YAML@stack{b}
\def\YAML@cdr#1#2\@nil{\ifx\relax#2\relax b\else#2\fi}
\newcommand\YAMLpush[1]{\xdef\YAML@stack{#1\YAML@stack}}
\newcommand\YAMLpop{\xdef\YAML@stack{\expandafter\YAML@cdr\YAML@stack\@nil}}
\newcommand\YAMLcomma{%
    \if b\expandafter\@car\YAML@stack\@nil ,%
    \else\YAMLpunct{,}\if k\expandafter\@car\YAML@stack\@nil\YAMLkeystyle\else\YAMLvaluestyle\fi\fi}
\lst@AddToHook{Init}{\gdef\YAML@stack{b}}
\makeatother
\lstdefinelanguage{YAML}{
    morecomment=[l]\#,
    morestring=[b]',
    morestring=[b]",
    stringstyle=\YAMLvaluestyle,
    commentstyle=\color{yaml_cyan}\mdseries,
    basicstyle=\scriptsize\ttfamily\YAMLkeystyle,
    keepspaces=true,
    showspaces=false,
    showstringspaces=false,
    rulecolor=\color{yaml_frame},
    frame=single,
    frameround={t}{t}{t}{t},
    framexleftmargin=6mm,
    backgroundcolor=\color{yaml_bg},
    aboveskip=\smallskipamount,
    belowskip=\smallskipamount,
    moredelim=**[il][\YAMLpunct{:}\YAMLvaluestyle\ ]{:\ },         
    moredelim=**[il][\YAMLcomma]{,},                               
    moredelim=**[il][\YAMLpunct{\char123}\YAMLpush{k}\YAMLkeystyle]{\{}, 
    moredelim=**[il][\YAMLpunct{[}\YAMLpush{v}\YAMLvaluestyle]{[}, 
    literate={\}}{{\YAMLpunct{\char125}\YAMLpop}}1
             {]}{{\YAMLpunct{]}\YAMLpop}}1
             {-\ }{{{\YAMLpunctstyle-}\ }}2
             {\ }{{\copyablespace}}1,
}

\DeclareFloatingEnvironment[
    fileext=lol,%
    placement=tbhp,%
    name=Listing
]{lstfloat}
\crefname{lstfloat}{listing}{listings}
\Crefname{lstfloat}{Listing}{Listings}
\crefname{appendix}{Appendix}{Appendices}
\Crefname{appendix}{Appendix}{Appendices}

\newacro{BSM}[BSM]{Beyond the Standard Model}

\newacro{CKM}[CKM]{Cabibbo--Kobayashi--Maskawa}

\newacro{KDE}[KDE]{kernel density estimate}

\newacro{PDF}[PDF]{probability density function}
\newcommand{\PDF}{\ac{PDF}\xspace}
\newacro{SM}[SM]{Standard Model}
\newcommand{\SM}{\ac{SM}\xspace}
\newacro{WET}[WET]{Weak Effective Theory}
\newcommand{\WET}{\ac{WET}\xspace}
\newcommand{\cpp}{\texttt{C++}\xspace}
\newcommand{\dynesty}{\texttt{dynesty}\xspace}
\newcommand{\EOS}{\texttt{EOS}\xspace}

\newcommand{\Jupyter}{\texttt{Jupyter}\xspace}
\newcommand{\matplotlib}{\texttt{matplotlib}\xspace}

\newcommand{\Python}{\texttt{Python}\xspace}
\newcommand{\pypmc}{\texttt{pypmc}\xspace}

\newcommand{\YAML}{\texttt{YAML}\xspace}
\newcommand{\code}[1]{\lstinline{#1}} 

\newcommand{\pyclass}[1]{\textcolor{black}{\lstinline{#1}\xspace}}

\newcommand{\pymodule}[1]{\textcolor{black}{\lstinline{#1}\xspace}}

\newcommand{\yamlkey}[1]{\texttt{\YAMLkeystyle #1}}
\newcommand{\yamlvalue}[1]{\texttt{\YAMLvaluestyle #1}}

\newcommand{\GeV}{\ensuremath{\mathrm{GeV}}}

\newcommand{\vecth}{\ensuremath{\vec\vartheta}}

\definecolor{remarkgray}{gray}{0.40}

\usepackage[most]{tcolorbox}

\newtcolorbox{remarkm}{
  breakable,
  enhanced,
  blank,
  left skip=8pt,
  colupper=remarkgray,
  borderline west={2pt}{-8pt}{lightgray}
}

\makeatletter
    \def\dvd{\@ifstar\@@dvd\@dvd}
    \newcommand{\@dvd}[1]{\textcolor{purple}{[\textbf{DvD:} #1]}}
    \newcommand{\@@dvd}[1]{\textcolor{purple}{#1}}
    \def\mr{\@ifstar\@@mr\@mr}
    \newcommand{\@mr}[1]{\textcolor{cyan}{[\textbf{MR:} #1]}}
    \newcommand{\@@mr}[1]{\textcolor{cyan}{#1}}
\makeatother

\journalname{Eur. Phys. J. C}

\begin{document}

\newcommand{\EOSversion}{1.1\xspace}

\title{%
    \includegraphics[width=0.15\linewidth]{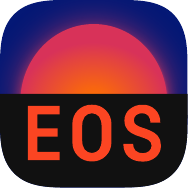}
    \hspace*{0.05\linewidth}
    \parbox[b][0.15\linewidth][c]{0.75\linewidth}{EOS version \EOSversion \\ A Software for Flavor Physics Phenomenology}\linebreak
}
\dedication{EOS-2026-04,IPPP/26/69,MITP-26-034,ZU-TH 34/26}
\author{%
    The EOS Authors\thanksref{v1.1} ---
    N.~C.~Balz~\orcidlink{0009-0008-3966-8450}\thanksref{aff:BonnHISKP} \and
    C.~Bolognani~\orcidlink{0000-0003-3752-6789}\thanksref{aff:TUDO} \and
    M.~Burgos~Marcos~\orcidlink{0009-0001-9716-0793}\thanksref{aff:Maastricht,aff:NIKHEF} \and
    C.~Earnshaw~\orcidlink{0009-0004-7653-3943}\thanksref{aff:IPPP} \and
    L.~G\"artner~\orcidlink{0000-0002-3643-4543}\thanksref{aff:LMU} \and
    N.~Gubernari~\orcidlink{0000-0002-2278-2707}\thanksref{aff:BonnHISKP} \and
    F.~Herren~\orcidlink{0000-0003-2810-1420}\thanksref{aff:Zuerich} \and
    M.~Kirk~\orcidlink{0000-0003-0845-7227}\thanksref{aff:IPPP} \and
    S.~K\"urten\thanksref{aff:TUM} \and
    V.~Kuschke~\orcidlink{0009-0005-7923-4006}\thanksref{aff:PSI,aff:ZuerichAST} \and
    D.~Leljak\thanksref{aff:RBI} \and
    P.~L\"ughausen~\orcidlink{0000-0002-2813-4620}\thanksref{aff:Siegen} \and
    E.~McPartland\thanksref{aff:IPPP} \and
    S.~Meiser~\orcidlink{0000-0002-9985-8095}\thanksref{aff:UB} \and
    S.~Mutke~\orcidlink{0009-0002-0509-7932}\thanksref{aff:BonnHISKP} \and
    F.~Novak\thanksref{aff:TUM} \and
    M.~Reboud~\orcidlink{0000-0001-6033-3606}\thanksref{aff:Orsay} \and
    M.~E.~Smith~\orcidlink{0000-0002-3872-1917}\thanksref{aff:Imperial} \and
    D.~Suelmann~\orcidlink{0009-0009-3986-0909}\thanksref{aff:TUDO} \and
    D.~van~Dyk~\orcidlink{0000-0002-7668-810X}\thanksref{aff:IPPP} \and
    K.~K.~Vos~\orcidlink{0000-0002-4258-4062}\thanksref{aff:Maastricht,aff:NIKHEF}%
}
\institute{%
\setlength{\parindent}{0pt}
\label{aff:BonnHISKP}
Helmholtz-Institut f\"ur Strahlen- und Kernphysik (Theorie), Bethe Center for Theoretical Physics, and Cluster of Excellence ``Color meets Flavor'',
Universit\"at Bonn, D-53115 Bonn, Germany\and
\label{aff:TUDO}
Department of Physics and Cluster of Excellence ``Color meets Flavor'', TU Dortmund University, D-44221 Dortmund, Germany\and
\label{aff:Maastricht}
Gravitational Waves and Fundamental Physics (GWFP),
Maastricht University, NL-6229 GT Maastricht, the Netherlands\and
\label{aff:NIKHEF}
Nikhef, NL-1098 XG Amsterdam, the Netherlands\and
\label{aff:IPPP}
Institute for Particle Physics Phenomenology and Department of Physics, Durham University, Durham DH1 3LE, UK\and
\label{aff:LMU}
Ludwig-Maximilians-Universität München, D-85748 Garching b. M\"unchen, Germany\and
\label{aff:Zuerich}
Physik-Institut, Universit\"at Z\"urich, Winterthurerstrasse 190, CH-8057 Z\"urich, Switzerland\and
\label{aff:TUM}
Physik Department T31, Technische Universit\"at M\"unchen, D-85748 Garching, Germany\and
\label{aff:PSI}
Paul Scherrer Institut, CH-5232 Villigen, Switzerland\and
\label{aff:ZuerichAST}
Institut f\"ur Astrophysik, Universit\"at Z\"urich, Winterthurerstrasse 190, CH-8057 Z\"urich, Switzerland\and
\label{aff:RBI}
Rudjer Boskovic Institute, Division of Theoretical Physics, Bijeni\v cka 54, HR-10000 Zagreb, Croatia\and
\label{aff:Siegen}
Theoretische Physik 1, Naturwissenschaftlich-Technische Fakult\"at, Universit\"at Siegen, D-57068 Siegen, Germany\and
\label{aff:UB}
Departament de Física Quàntica i Astrofísica and ICCUB,
Universitat de Barcelona, E08028 Barcelona, Catalunya\and
\label{aff:Orsay}
IJCLab, Pôle Théorie, CNRS/IN2P3 et Université Paris-Saclay, F-91405 Orsay, France\and
\label{aff:Imperial}
Imperial College London, SW7 2AZ London, United Kingdom
}
\date{\today}
\thankstext{v1.1}{Authors of this paper have contributed changes to the software since \EOS version 1.0.}

\maketitle

\setlength{\parindent}{0pt}

\begin{abstract}
    \EOS is an open-source software package for a variety of computational tasks in flavor physics.
    In this article, we describe new developments for version 1.1 of this software and
    discuss changes with respect to version 1.0.
    These developments and changes include theory predictions for
    additional decay and scattering processes, such as semileptonic charm decays,
    hadronic tau decays, and $e^+e^- \to \text{hadrons}$ scattering.
    Overall, these changes have contributed 1063 additional
    (pseudo-)observables.
    Beyond the physics capabilities of the software, the changes include
    improving the quality and reducing the computational costs of posterior sampling;
    providing a more versatile and testable figure framework; and
    improving the overall user experience.
\end{abstract}

\setcounter{tocdepth}{3}
\tableofcontents

\bigskip

This article is not a user manual for \EOS. It describes the changes to the software since
version~1.0 and complements, rather than replaces, the description of the software's design and use
given in Ref.~\cite{EOSAuthors:2021xpv}. The user guide and the reference documentation of all
\Python and \cpp interfaces, analysis-file keys, and tasks are part of the online
documentation~\cite{EOS:doc}.
We first summarize the changes in \cref{sec:changes}, then describe the new theory predictions in
\cref{sec:new-observables}, the new analysis file format in \cref{sec:analysis-files}, the improvements
to mode finding and posterior sampling in \cref{sec:sampling}, and the new plotting framework in
\cref{sec:plotting}. We conclude in \cref{sec:summary}. The changes that break backward
compatibility with version~1.0 are collected in \cref{app:breaking-changes}.

%
%
%
%
\section{Overview of Changes}
\label{sec:changes}

In the five years since the release of version 1.0~\cite{EOSAuthors:2021xpv},
the \EOS software has undergone active development,
with contributions from 25 authors.
This development was directly connected to 21 peer-reviewed publications
\cite{%
Blake:2022vfl,Gubernari:2022hxn,Amhis:2022vcd,Leljak:2023gna,%
Gubernari:2023puw,Bolognani:2023mcf,Amhis:2023mpj,Hanhart:2023fud,%
Lughausen:2024pmk,Bolognani:2024cmr,Kirk:2024oyl,Meiser:2024zea,%
Herren:2025cwv,BurgosMarcos:2025xja,Bordone:2025jur,%
Melic:2025uha,Kirk:2025ajn,Earnshaw:2025dhd,%
Hoferichter:2026jlh,Bolognani:2026yxj,Gubernari:2026sqc%
}.
Moreover, \EOS has been used in 8 experimental analyses
by the Belle~II~\cite{Belle-II:2025lfq}, CMS~\cite{CMS:2024syx}, and LHCb~\cite{LHCb:2022vje,LHCb:2022qnv,LHCb:2023gel,LHCb:2024onj,LHCb:2025pxz,LHCb:2025mqb}
experiments.
User feedback and research-driven development have led to a substantial expansion of both
the physics-related and utility-related source code,
yielding a net growth of 180k lines of code. We summarize the major changes to the code here
and provide detailed descriptions of these changes in subsequent sections.
\begin{itemize}
    \item Since version 1.0, we have added a wide variety of theoretical predictions
    of processes relevant to flavor physics phenomenology, including
    nonleptonic decays of $B$ mesons, semileptonic decays of $D_{(s)}^{(*)}$
    mesons and $\Lambda_c$ baryons, two-body hadronic $\tau$ decays, and $e^+e^-\to \text{hadrons}$ production.
    We discuss these additions in detail in \cref{sec:new-observables}.
    \smallskip

    \item \EOS version 1.1 fully supports a versatile analysis file format, which is documented in \cref{sec:analysis-files}.
    This change ensures that analyses can be stored in a human-readable file
    and that the verification and re-analysis of previous studies are possible.
    Moreover, thanks to this analysis file format, as of version 1.1 \EOS supports the evaluation of external
    likelihoods expressed in the HistFactory format~\cite{Cranmer:2012sba}
    by using the \texttt{pyhf} software~\cite{Heinrich:2021gyp}.

    \begin{remarkm}
    For releases after version 1.1, we guarantee a stable interface: code written against
    the \EOS \Python and \cpp interfaces continues to work, unless a change is required to fix a software bug.
    We do not extend this guarantee to the names and definitions of parameters, constraints, and observables,
    which keep evolving with the physics; an analysis that refers to them may therefore require adaptation.
    The breaking changes introduced since version 1.0 are collected in \cref{app:breaking-changes}.
    \end{remarkm}

    \item Since version 1.0, the default sampling method has been switched to dynamic nested
    sampling~\cite{Higson:2018}, as provided by the \dynesty package~\cite{Speagle:2020,dynesty:v3.0.0}.
    Subsequently, the \pyclass{eos.Analysis} class has been rewritten. The rewrite improves sampling speed
    and sampling quality by sampling in the space of the cumulative prior probability.
    We quantify the latter changes and document all changes in \cref{sec:sampling}.

    \begin{remarkm}
    This rewrite leaves the interface of version 1.0 unchanged.
    \end{remarkm}

    \item \EOS version 1.1 includes a new plotting framework contained in the \Python module
    \texttt{eos.figure}. This framework logically separates figures, plots, and plot items from each other.
    The new framework is based on standard \matplotlib functionality; the rewrite now makes it possible
    to transparently use \EOS-specific functionality within a \matplotlib figure. Moreover, it
    facilitates testing the code. The new framework is described in \cref{sec:plotting}.

    \begin{remarkm}
    The previous plotting framework, \texttt{eos.plot}, was deprecated in version 1.0.21.
    It has been removed in version 1.1.
    \end{remarkm}

    \item In preparation for \EOS version 1.1, the form-factor parametrizations have been reorganized:
    first, parametrizations respecting the dispersive bounds of the form factors
    (\texttt{ABR2022}~\cite{Amhis:2022vcd}, \texttt{BFW2010}~\cite{Bharucha:2010im}, and \texttt{BMRvD2022}~\cite{Blake:2022vfl})
    have been renamed to Series Expansion (\texttt{SE}). Their functionality remains the same.
    They can now be used simultaneously in a single analysis with a common dispersive bound.
    The interaction with the dispersive bound makes this rather specialized; the \texttt{SE}
    parametrization is only available for mesonic $P\to P$, mesonic $P\to V$, baryonic
    $1/2^+\to 1/2^+$, and baryonic $1/2^+\to 3/2^-$ transitions.
    Hence, a new Simplified Series Expansion (\texttt{SSE}) has also been added, similar to
    the popular \texttt{BSZ2015} parametrization~\cite{Bharucha:2015bzk}.
    This parametrization is available for mesonic $P\to P$, mesonic $P\to V$, and baryonic $1/2^+\to 1/2^+$ transitions.
    Unlike for \texttt{SE}, \EOS provides default values for its parameters, obtained from fits to theoretical inputs for
    26 transitions~\cite{EOS-DATA-2026-04}; it has therefore been selected as the default parametrization.
    The distinction between full and simplified series expansions follows the spirit of the nomenclature
    proposed in Ref.~\cite{Bharucha:2010im}.
    Further specialized parametrizations remain available.
    \smallskip

    \item Starting with \EOS version 1.1, we fully support installation on macOS (Apple Silicon only).
    Binary wheels are provided for Linux on \texttt{x86-64} and \texttt{AArch64}, as before, and now also for
    macOS on Apple Silicon. On each of these platforms, \EOS is installed using
    \begin{lstlisting}[%
        language=bash
    ]
        pip install eoshep       # for a versioned release
        pip install --pre eoshep # for the most recent weekly build
\end{lstlisting}
    \bigskip

    \item Beyond the \Python wheels, \EOS is now also distributed as a container image, which bundles a
    fixed \EOS release with all of its dependencies. The image is published for the same two
    Linux architectures as the wheels and is intended for analyses that must remain reproducible
    independently of the software installed on the host. Although the image itself is a Linux image,
    it can also be run on macOS through Apple's \texttt{container} tool and on Windows through the
    Windows Subsystem for Linux. It is obtained using
    \begin{lstlisting}[%
        language=bash
    ]
        docker pull eoshep/eos:latest # for the latest version or
        docker pull eoshep/eos:1.1    # tagged for a specific release
\end{lstlisting}
\end{itemize}

%
%
%
%
\section{Expanding the Coverage of Flavor Observables}
\label{sec:new-observables}

\begin{figure}
    \centering
    \includegraphics[width=0.41\linewidth]{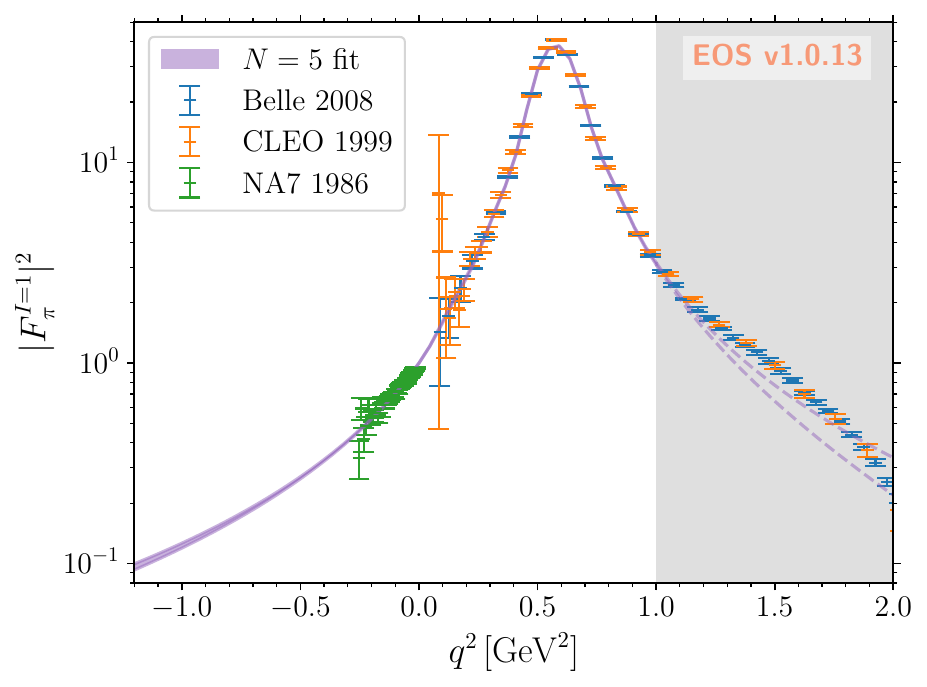} \hfill
    \includegraphics[width=0.55\linewidth]{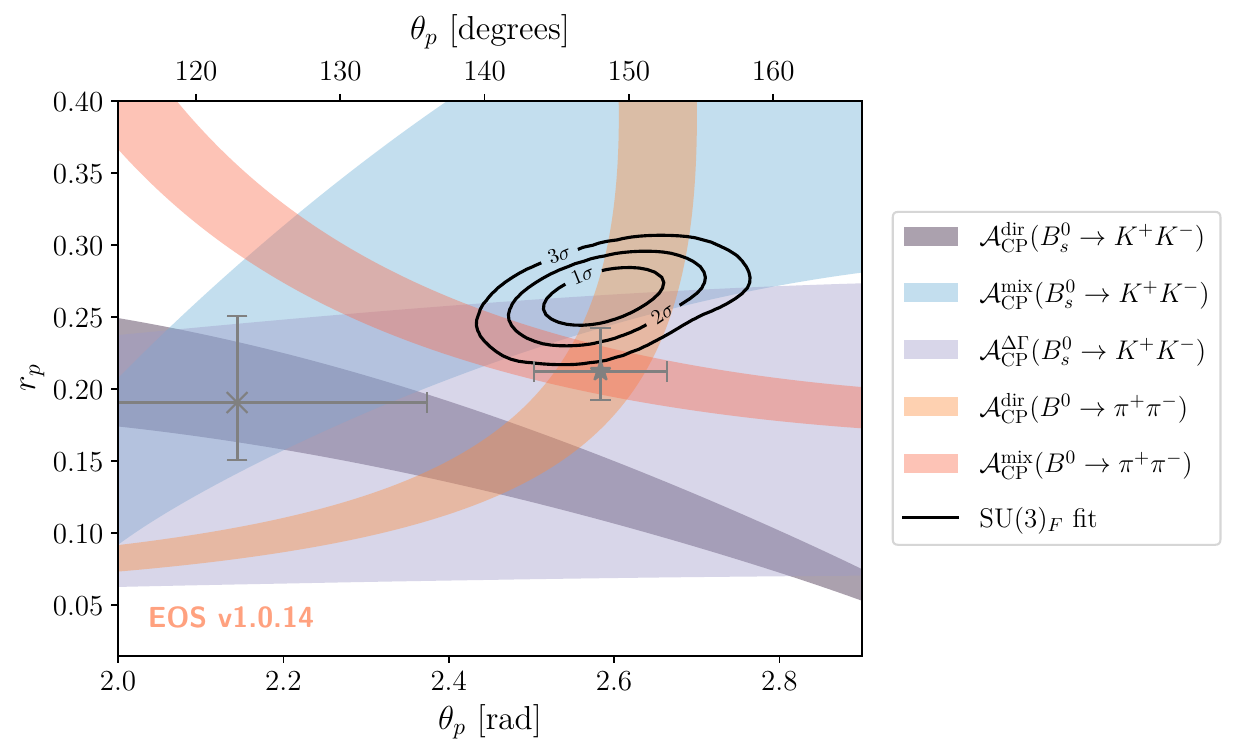}
    \caption{%
        Representative examples of published figures produced in the course of physics analyses.
        The left panel shows a fit of the pion vector form factor to data extracted from hadronic $\tau$ decays,
        using a simple form-factor parametrization proposed in Ref.~\cite{Kirk:2024oyl}.
        The right panel illustrates the tension in the penguin-to-tree ratio between the experimental
        constraints from $B^0\to\pi^+\pi^-$ and $B_s^0\to K^+K^-$ when an $\mathrm{SU}(3)$ symmetry is assumed for the light-quark flavors,
        as produced for Ref.~\cite{BurgosMarcos:2025xja}.
    }
    \label{fig:representative-figures}
\end{figure}

New flavor observables are frequently added to EOS as part of a planned physics analysis.
\Cref{fig:representative-figures} shows two figures produced in the course of such analyses.
Alongside the direct implementation of the observables, the relevant effective-theory operators and coefficients must be defined, together with experimental or theoretical constraints.
Since version 1.0, we have added the following six sectors of the \WET:
\begin{itemize}
    \item $\bar{s}b\bar{c}u$ and $\bar{d}b\bar{c}u$,
    \item $\bar{s}c\bar{\nu}\ell$ and $\bar{d}c\bar{\nu}\ell$,
    \item $c\to u\left\lbrace \bar{q}q,\, \gamma,\, \ell^+\ell^-\right\rbrace$,
    \item $\bar{u}s\bar{\ell}\nu$.
\end{itemize}
Over the same period, the number of constraints has grown from 407 to 732:
we have added 330 constraints, of which 257 are experimental and 73 are theoretical,
renamed 17, and removed 5; see \cref{app:breaking-changes}.
In the following, we summarize the major additions and changes to
the set of flavor observables in \EOS, organized by transition type.

\paragraph{(Semi)leptonic $b$-quark decays}
Semileptonic $b$ decay observables have been part of the \EOS
software since before version 1.0. Here, we list
the major additions and changes since the release of version 1.0.
\begin{itemize}
  \item Additions include branching ratios and angular observables
        for the photoleptonic decay $B^-\to \gamma\ell^-\bar\nu$
        and for the closely related four-lepton decay
        mode $B^-\to \ell^-\bar\nu\ell^{\prime +}\ell^{\prime -}$.
        As part of the latter, we implemented the full set of
        $B^-\to \gamma^*$ form factors in a basis free of
        kinematic singularities~\cite{Kurten:2022zuy}.
  \item The code for semileptonic $\bar{B}_q\to D_q^{(*)}\ell^-\bar\nu$
        decays has been unified with that for other
        $\bar{B}\to P\ell^-\bar\nu$ and $\bar{B}\to V\ell^-\bar\nu$
        modes.
        The decay modes $\bar{B}_c\to J/\psi\,\ell^-\bar\nu$,
        $\bar{B}_q\to\eta\ell^-\bar\nu$, and $\bar{B}_q\to\eta'\ell^-\bar\nu$ have been added~\cite{Bolognani:2026yxj,Melic:2025uha}.
  \item The form factors for $B^-\to\pi^+\pi^-\ell^-\bar\nu$
        have been extended with a parametrization that respects
        dispersive bounds and uses a double series expansion~\cite{Herren:2025cwv}. The $\pi\pi$ lineshapes are implemented through model-independent Omnès functions~\cite{Omnes:1958hv} with scattering input from Refs.~\cite{Garcia-Martin:2011iqs,Colangelo:2018mtw,Daub:2015xja,Ropertz:2018stk}.
  \item The branching ratios of $\bar{B}\to P\ell^-\bar\nu$
        and $\bar{B}\to V\ell^-\bar\nu$ decays, differential in and integrated over the
        momentum component $k_\perp$, have been added in support of Ref.~\cite{Earnshaw:2025dhd}.
        This has made it possible to exploit public LHCb data sets
        that make use of this momentum component~\cite{LHCb:2020cyw}.
\end{itemize}
\begin{remarkm}
In total, these changes have added 163 observables.
\end{remarkm}

\paragraph{Rare and radiative $b$-quark decays}
Rare and radiative $b$ decay observables have been a mainstay of the
\EOS software since its first commits. As a consequence, they have
accumulated considerable technical debt, which has led to a large number of breaking changes;
see also \cref{app:breaking-changes}.
The major additions and changes include:
\begin{itemize}
  \item Exclusive $b\to s\,\ell^+\ell^-$
        observables have been completely rewritten, covering
        $\bar{B}\to K^{(*)}\ell^+\ell^-$ and
        $\Lambda_b\to \Lambda(\to p \pi)\ell^+\ell^-$ decays.
        The rewrite makes it possible to change the predictions
        for the nonlocal contributions at run time.
        In the course of these changes, the decays
        $\bar{B}_s\to\phi\ell^+\ell^-$
        and $\Lambda_b\to\Lambda(1520)\ell^+\ell^-$ have been added.
        As a side effect of implementing the dispersively bounded
        nonlocal form factors~\cite{Gubernari:2022hxn},
        \EOS now also supports predictions
        for the decays $\bar{B}\to \bar{K}^{(*)}\psi$
        and $\bar{B}_s\to\phi\psi$, where $\psi$ is either $J/\psi$
        or $\psi(2S)$.

  \item In the same spirit, the observables of exclusive
        $\bar{B}\to \bar{K}^*\gamma$ decays have been rewritten,
        and the decay $\Lambda_b\to\Lambda(1520)\gamma$ has been added.

  \item In support of Refs.~\cite{Amhis:2023mpj,Gartner:2024muk,Melic:2025uha}, \EOS has been
        extended with predictions for exclusive $b\to s\,\nu\bar\nu$ decays,
        including $\bar{B}\to \bar{K}^{(*)}\nu\bar\nu$, $\bar{B}_s\to\phi\nu\bar\nu$,
        $\Lambda_b\to\Lambda\nu\bar\nu$, $\bar{B}_s\to\eta\nu\bar\nu$, and
        $\bar{B}_s\to\eta'\nu\bar\nu$.
\end{itemize}
\begin{remarkm}
In total, these changes have added 142 observables.
\end{remarkm}

\paragraph{Nonleptonic $b$-quark decays}
Nonleptonic $b$ decays have been added since the \EOS v1.0 release,
following two distinct approaches.
\begin{itemize}
  \item Charmless nonleptonic $B\to PP$ decays have been implemented
        within the frameworks of $\mathrm{SU}(3)_\mathrm{F}$ symmetry
        and of topological amplitudes, in support of Ref.~\cite{BurgosMarcos:2025xja}.
        For both frameworks, \EOS covers
        the full set of $\bar{B}^0$, $B^-$, and $\bar{B}_s^0$ two-body decays
        to $\pi\pi$, $K\bar K$, $K\pi$, $\bar{K}\pi$, $\eta^{(\prime)}\pi$,
        $\eta^{(\prime)}K$, $\eta^{(\prime)}\bar{K}$, $\eta^{(\prime)}\eta^{(\prime)}$,
        and $K_S\pi^0$. In total, $150$ new observables
        have been implemented, including branching ratios and CP asymmetries.
  \item Nonleptonic $b\to c \bar{u} q$ decays, with $q=d,s$, have been added in
        support of Ref.~\cite{Meiser:2024zea}, within the framework of
        QCD factorization. Here, the \EOS observables
        cover the decays
        $\bar{B}^0\to D^{(*)+}K^-$ and $\bar{B}_s^0\to D_s^{(*)+}\pi^-$, i.e., class-I
        $\bar{B}$ decays without contributions from annihilation topologies.
\end{itemize}
\begin{remarkm}
In total, these changes have added $154$ observables, alongside $6$ pseudo-observables
for the $B\to PP$ amplitudes.
\end{remarkm}

\paragraph{(Semi)leptonic and rare $c$-quark decays}
Leptonic, semileptonic, and rare $c$ decays were added after the \EOS v1.0 release.
They have since been expanded towards the feature set available for semileptonic $b$ decays.
The decay processes include:
\begin{itemize}
  \item leptonic decays $D_{(s)}^+\to\ell^+\nu$ and $D_s^{*+}\to\ell^+\nu$;
  
  \item semileptonic decays of a $D$ meson such as $D\to K\ell^+\nu$, $D_{(s)}^+\to\eta\ell^+\nu$,
        and $D^+_{(s)}\to\eta'\ell^+\nu$;

  \item semileptonic decays of a $\Lambda_c$ baryon such as
        $\Lambda_c^+\to\Lambda^0\ell^+\nu$ and $\Lambda_c^+\to n\,\ell^+\nu$;
  \item and the rare decay $\Lambda_c^+\to p\,\ell^+\ell^-$, which proceeds through a
        flavor-changing neutral current.
\end{itemize}
Most of these changes have been made in support of Refs.~\cite{Bolognani:2024cmr,thesisSuelmann}.
\begin{remarkm}
In total, these changes have added 37 observables.
\end{remarkm}

\paragraph{Hadronic $\tau$-lepton decays}
Hadronic $\tau$ decays are new to \EOS since the v1.0 release.
The decay processes are presently limited to
\begin{itemize}
  \item decays to a single hadron, $\tau^-\to K^-\nu$;
  \item decays to a pair of hadrons, $\tau^-\to [K\pi]^-\nu$.
\end{itemize}
All of these changes have been made in support of Ref.~\cite{Kirk:2025ajn}.
\begin{remarkm}
In total, these changes have added 9 observables.
\end{remarkm}

\paragraph{Scattering processes}
Scattering processes are another addition following the \EOS v1.0 release.
Electron-positron scattering in the charmonium region has been implemented using a coupled-channel
$K$-matrix framework, which is applicable in the energy region close to the open-charm threshold.
The framework couples the elastic $e^+e^-$ channel, the exclusive $D^+D^-$ and $D^0\bar D^0$ channels,
and a set of effective channels that collect the non-$c\bar{c}$ final states. Supported observables are
the corresponding cross sections, the inclusive ratio $R$, and the widths and spectral functions of the
resonances, which are the $J/\psi$, the $\psi(2S)$, and the $\psi(3770)$.
These changes support the analysis of Ref.~\cite{Hanhart:2023fud}.
\begin{remarkm}
In total, these changes have added 31 observables.
\end{remarkm}

\paragraph{(Semi)leptonic $s$-quark decays}
Leptonic and semileptonic $s$ decays were likewise added after the \EOS v1.0 release.
The decay processes include:
\begin{itemize}
  \item leptonic decays $K^-\to\ell^-\bar\nu$ (also known as $K_{\ell2}$ decays);
  \item and semileptonic decays $\bar{K}\to \pi\ell^-\bar\nu$ (also known as $K_{\ell3}$ decays).
\end{itemize}
Like the hadronic $\tau$ decays, these have been added in support of Ref.~\cite{Kirk:2025ajn}.
\begin{remarkm}
In total, these changes have added 4 observables.
\end{remarkm}

\paragraph{(Nonlocal) form factors}
The form factors have been extended alongside the observables above.
Additionally, an implementation of the pion isovector vector form factor, $F_\pi$, has been added in support of Ref.~\cite{Kirk:2024oyl}.
\begin{remarkm}
In total, these changes have added 395 pseudo-observables for the local form factors,
and a further 122 for the nonlocal form factors.
\end{remarkm}

%
%
%
%
\section{Organizing an Analysis in a Single File}
\label{sec:analysis-files}

The \EOS user guide~\cite{EOS:doc} illustrates the various \EOS use cases and covers relatively simple phenomenological analyses.
For these representative examples, the number of observables, parameters, and statistical constraints is rather small compared with
that of full phenomenological studies, e.g., Ref.~\cite{Bolognani:2024cmr}.
Typical analyses carried out using \EOS require a high degree of organization, especially to reduce
the amount of repeated information.
To this end, we have modified \EOS to support the use of a so-called analysis file, which provides the means to store
the description of one or more posteriors and their associated data within a \YAML file.
This is particularly beneficial when the posteriors in the analysis share common elements. In this
section, we illustrate the setup with a real-world example: the analysis of Ref.~\cite{Bolognani:2024cmr},
whose \EOS analysis file is available from its supplementary material~\cite{EOS-DATA-2024-01}.
That file was written for \EOS version 1.0; the listings shown below, and the analysis file accompanying
this article, have been adapted to version 1.1. In particular, the form-factor parametrizations named
\texttt{BFW2010} and \texttt{BMRvD2022} in Ref.~\cite{Bolognani:2024cmr} have both been renamed to \texttt{SE} in version 1.1.
This analysis of exclusive $c\to s\ell^+\nu_\ell$ transitions combines the leptonic $D_s^{(*)}\to \ell^+\nu_\ell$ decays
with the semileptonic $D\to K\ell^+\nu_\ell$ and $\Lambda_c\to\Lambda\ell^+\nu_\ell$ decays.
It involves a number of different posteriors that share some or all of their
hadronic nuisance parameters; such sharing of nuisance parameters is a common occurrence in flavor physics analyses.

The analysis file format is designed to ensure a lean description of the analysis and its posteriors:
the description of parameters shared among several posteriors should not be repeated, since repetition is liable to introduce
inconsisten\-cies or even errors. To this end, shared parameter descriptions are grouped into named \emph{prior components}.
For the present example, parameters
describing the hadronic matrix elements of semileptonic $D\to K$ transitions appear not only in posteriors that
exclusively use $D\to K\ell^+\nu_\ell$ data (e.g., \yamlvalue{CKM-D-to-K-l-nu}) but also in ``global'' posteriors that use all available
exclusive $c\to s\ell^+\nu_\ell$ data (e.g., \yamlvalue{CKM-all}).
Hence, the $D\to K$-specific parameters are grouped into a dedicated prior component named \yamlvalue{FF-D-to-K-order3}.

Similarly, the experimental data on the $D\to K\ell^+\nu_\ell$ branching ratios
are used within the likelihood of both the $D\to K$-specific posteriors and the global posteriors.
It is described in named \emph{likelihood components} such as \yamlvalue{EXP-D0-to-K-e-nu}, which are then referenced in the
description of the relevant posterior(s).

In the following, we describe the format in detail using this example.

Every \EOS analysis file is a \YAML file that describes one or more Bayesian posteriors and associated data.
At the top level, the format includes the following \YAML keys:
\begin{description}
    \item[\yamlkey{format\_version}] The version of the analysis file format (optional; defaults to $1$, the version predating format versioning). \EOS refuses to load a file that declares a version newer than the one it supports.
    \item[\yamlkey{priors}] The list of prior components available within the analysis (mandatory).
    See \cref{sec:analysis-files:priors} for an introduction to defining prior components.
    \item[\yamlkey{likelihoods}] The list of likelihood components within the analysis (optional; \EOS warns if it is absent).
    See \cref{sec:analysis-files:likelihoods} for an introduction to defining likelihood components.
    \item[\yamlkey{posteriors}] The list of posteriors within the analysis (mandatory).
    See \cref{sec:analysis-files:posteriors} for an introduction to defining posteriors.
    \item[\yamlkey{predictions}] The list of posterior predictions performed as part of the analysis (optional).
    See \cref{sec:analysis-files:predictions} for an introduction to defining posterior predictions.
    \item[\yamlkey{masks}] The list of masks for filtering posterior samples (optional).
    See \cref{sec:analysis-files:masks} for an introduction to defining masks.
    \item[\yamlkey{parameters}] The list of custom parameters defined within the scope of the analysis (optional).
   See \cref{sec:analysis-files:parameters} for an introduction to defining custom parameters.
    \item[\yamlkey{observables}] The list of custom observables defined within the scope of the analysis (optional).
    See \cref{sec:analysis-files:observables} for an introduction to defining custom observables.
    \item[\yamlkey{constraints}] The list of custom constraints that are forward-declared within the analysis (optional).
    See \cref{sec:analysis-files:constraints} for an introduction to declaring custom constraints.
    \item[\yamlkey{steps}] The list of steps to reproduce the analysis (optional).
    See \cref{sec:analysis-files:steps} for an introduction to defining the analysis steps.
    \item[\yamlkey{metadata}] The bibliographic information about the analysis (optional).
    See \cref{sec:analysis-files:metadata} for an introduction to providing the metadata.
    \item[\yamlkey{figures}] The list of figures to be produced from the analysis (optional).
    See \cref{sec:plotting} for an in-depth description of the format.
\end{description}
An \EOS analysis can be loaded from a file via the \pyclass{eos.AnalysisFile} class.
Note that loading an analysis file automatically registers all custom observables and parameters defined in it.
We refer to the \EOS documentation on the Analysis File Format~\cite[\href{https://eoshep.org/doc/releases/v1.1/reference/analysis-file-format.html}{Reference Manual/Analysis File Format}]{EOS:doc} for a comprehensive description of the syntax of each \YAML key.

Beyond describing an analysis, \EOS provides tools to check an analysis file and to summarize its results.
The \code{validate} task detects mistakes in an analysis file before any time-consuming computation is carried out;
see \cref{sec:analysis-files:validation}.
The \code{report} task turns the results of an analysis into a human-readable document;
see \cref{sec:analysis-files:reporting}.

\subsection{Priors}
\label{sec:analysis-files:priors}

The \yamlkey{priors} key contains a list of \emph{named} prior components, each represented by a map.
Each prior component includes two mandatory keys:
\begin{description}
  \item[\yamlkey{name}] The file-unique name of this prior component (mandatory).
  \item[\yamlkey{descriptions}] The ordered list of parameter descriptions used to construct this prior component (mandatory).
\end{description}
For a univariate prior, \yamlkey{descriptions} contains only a single parameter description,
which is a key/value map describing the parameter's prior.
For an uncorrelated multivariate prior, \yamlkey{descriptions} contains more than one parameter description.
The format for a parameter description is the same as that of the \yamlkey{priors} keyword argument
of the \pyclass{eos.Analysis} constructor.
For example, the following listing defines two named prior components \yamlvalue{CKM} (univariate) and \yamlvalue{FF-D-to-K-order3} (multivariate but uncorrelated).

\begin{lstlisting}[language=YAML]
priors:
  - name: CKM
    descriptions:
      - { parameter: 'CKM::abs(V_cs)',   min: 0.88,    max: 1.03,   type: 'uniform' }

  - name: FF-D-to-K-order3
    descriptions:
      - { parameter: 'D->K::a^f+_0@SE',  min:   0.04219, max:   0.04367, type: 'uniform' }
      - { parameter: 'D->K::a^f+_1@SE',  min:  -0.10296, max:  -0.07851, type: 'uniform' }
      - { parameter: 'D->K::a^f+_2@SE',  min:  -0.09755, max:   0.06022, type: 'uniform' }
      - { parameter: 'D->K::a^f+_3@SE',  min:  -0.10838, max:   0.89484, type: 'uniform' }
      - { parameter: 'D->K::a^f0_1@SE',  min:  -0.25136, max:  -0.23142, type: 'uniform' }
      - { parameter: 'D->K::a^f0_2@SE',  min:   0.00756, max:   0.27770, type: 'uniform' }
      - { parameter: 'D->K::a^f0_3@SE',  min:  -0.23736, max:   1.     , type: 'uniform' }
      - { parameter: 'D->K::a^fT_0@SE',  min:   0.07963, max:   0.09720, type: 'uniform' }
      - { parameter: 'D->K::a^fT_1@SE',  min:  -0.18138, max:  -0.04286, type: 'uniform' }
      - { parameter: 'D->K::a^fT_2@SE',  min:  -0.64666, max:   0.39906, type: 'uniform' }
      - { parameter: 'D->K::a^fT_3@SE',  min:  -1.     , max:   1.     , type: 'uniform' }
\end{lstlisting}

For a correlated multivariate prior component, \yamlkey{descriptions} contains a single parameter description.
This description must feature the key \yamlkey{type} with the value \yamlvalue{constraint}, and the key
\yamlkey{constraint}, whose value is the qualified name of an existing \EOS constraint.
Although not part of this analysis example, the following listing illustrates the organization of a correlated multivariate prior.

\begin{lstlisting}[%
    language=YAML%
]
priors:
  - name: FF-rho
    descriptions:
      - { type: 'constraint', constraint: 'B->rho::FormFactors[parametric,LCSR]@BSZ:2015A' }
\end{lstlisting}

\subsection{Likelihoods}
\label{sec:analysis-files:likelihoods}

The \yamlkey{likelihoods} key contains a list of named likelihood components. Each likelihood component has a mandatory \yamlkey{name}
and gathers its contributions through at least one of the keys \yamlkey{constraints}, \yamlkey{manual\_constraints}, and \yamlkey{pyhf}:
\begin{description}
  \item[\yamlkey{name}] The unique name of this likelihood component (mandatory).
  \item[\yamlkey{constraints}] An ordered list of built-in \EOS constraints, referenced by their qualified names, that comprise this likelihood component (optional).
  \item[\yamlkey{manual\_constraints}] A key/value map of constraints defined inline within the analysis file, for data that is not part of the built-in \EOS database (optional).
  \item[\yamlkey{pyhf}] A likelihood imported from a \texttt{pyhf} workspace, i.e., a HistFactory statistical model (optional).
\end{description}
At least one of \yamlkey{constraints}, \yamlkey{manual\_constraints}, and \yamlkey{pyhf} must be present.

The following example illustrates a likelihood component consisting of a single built-in \yamlkey{constraints} entry.

\begin{lstlisting}[language=YAML]
likelihoods:
  - name: EXP-Dsstar-to-e-nu
    constraints:
      - 'D_s^*->e^+nu::BR@BESIII:2023C'
\end{lstlisting}

Besides referencing the built-in constraints, a likelihood component can define its own constraints inline via the \yamlkey{manual\_constraints} key.
Each entry is a key/value map whose key is a file-unique name for the constraint and whose value describes it. The key \yamlkey{type} selects the kind of constraint; for a \yamlvalue{Gaussian} constraint, the description provides the \yamlkey{observable} to be constrained
together with its \yamlkey{kinematics} and \yamlkey{options}, the measured central value \yamlkey{mean}, and its statistical and systematic uncertainties
\yamlkey{sigma-stat} and \yamlkey{sigma-sys} (each split into an upper and a lower part \yamlkey{hi}/\yamlkey{lo}). A \yamlkey{manual\_constraints} entry accepts the same values for \yamlkey{type} as built-in constraints.
The following example encodes a single measurement of the $D_s^+\to\mu^+\nu$ branching ratio as a Gaussian likelihood.
\begin{lstlisting}[language=YAML]
likelihoods:
  - name: EXP-Ds-to-mu-nu
    manual_constraints:
      D_s^+->mu^+nu::BR@BRvDV2024:
        type: Gaussian
        observable: D_s->lnu::BR
        kinematics: { }
        options: { l: mu }
        mean: 5.35e-03
        sigma-stat: { hi: 0.10e-03, lo: 0.10e-03 }
        sigma-sys: { hi: 0.0, lo: 0.0 }
        dof: 1
\end{lstlisting}
The constraint description format is identical to the one used for the built-in constraints database.
A description of this format is part of the \EOS documentation~\cite[\href{https://eoshep.org/doc/releases/v1.1/reference/constraint-format.html}{Reference Manual/Constraint Format}]{EOS:doc}.

An external likelihood can be imported through a \yamlkey{pyhf} key.
It requires a \yamlkey{file} key, which gives the path to the JSON file specifying the \texttt{pyhf} workspace with the HistFactory statistical model.
The parameters of the \texttt{pyhf} model can be identified with an \EOS observable or parameter through an optional \yamlkey{parameter\_map} key;
the observable can be either a built-in one or a custom one defined within the same analysis file; see \cref{sec:analysis-files:observables}.
Although not part of this analysis example, the following listing illustrates the organization of an external likelihood entry;
it is taken from the \code{examples/pyhf_corr.yaml} analysis file that ships with the \EOS source code~\cite{EOS:repo}.

\begin{lstlisting}[language=YAML]
likelihoods:
  - name: EXP-pyhf
    pyhf:
      file: workspace_corr.json
      parameter_map:
        mu: { name: 'B->pilnu::mu', kinematics: { q2_min: 1.0e-7, q2_max: 25.0 } }
\end{lstlisting}
The observable \yamlvalue{B->pilnu::mu} is of the latter kind: that same file defines it as the signal strength, i.e., the branching ratio of $\bar{B}\to\pi\ell^-\bar\nu$ in the \WET normalized to its \SM value.

\subsection{Posteriors}
\label{sec:analysis-files:posteriors}

The \yamlkey{posteriors} key contains a list of named posteriors. Each posterior contains three mandatory and two optional keys:
\begin{description}
    \item[\yamlkey{name}] The unique name of this posterior (mandatory).
    \item[\yamlkey{prior}] The list of named prior components that are used as part of this posterior (mandatory).
    \item[\yamlkey{likelihood}] The list of named likelihood components that are used as part of this posterior (mandatory).
    \item[\yamlkey{global\_options}] A key/value map providing global options, i.e., options that apply to all observables used within this posterior (optional).
    \item[\yamlkey{fixed\_parameters}] A key/value map providing values that override the \EOS-wide defaults of selected parameters (optional).
\end{description}

The following example illustrates the organization of two posteriors that share their hadronic nuisance prior and likelihood components
but use different physics models, namely \yamlvalue{CKM} and \yamlvalue{WET}.

\begin{lstlisting}[language=YAML]
posteriors:
  - name: CKM-all
    global_options:
      model: CKM
    fixed_parameters:
      'D->K::tp@SE': 5.56488
    prior:
      - CKM
      - DC-Ds-Dsstar
      - FF-D-to-K-order3
      - FF-Lc-to-L-order2
    likelihood:
      - TH-Ds-Dsstar-LQCD
      # components removed for brevity
      - EXP-Lc-to-L-mu-nu

  - name: WET-all
    global_options:
      model: WET
    fixed_parameters:
      'D->K::tp@SE': 5.56488
      'CKM::abs(V_cs)': 0.975
    prior:
      - WET-Re
      - WET-Im
      - DC-Ds-Dsstar
      - FF-D-to-K-order3
      - FF-Lc-to-L-order2
    likelihood:
      - TH-Ds-Dsstar-LQCD
      # components removed for brevity
      - EXP-Lc-to-L-mu-nu
\end{lstlisting}

\subsection{Predictions}
\label{sec:analysis-files:predictions}

The \yamlkey{predictions} key contains a list of named posterior predictions.
Each prediction evaluates a set of observables on the importance samples of a posterior, thereby propagating
the inferred uncertainties to the predicted quantities; producing such predictions is typically the last step of an analysis.
Each prediction has two mandatory and two optional keys:
\begin{description}
    \item[\yamlkey{name}] The unique name of this prediction (mandatory).
    \item[\yamlkey{observables}] The list of observables to be predicted (mandatory).
    \item[\yamlkey{global\_options}] A key/value map of options applied to all observables of this prediction (optional).
    \item[\yamlkey{fixed\_parameters}] A key/value map of parameters held fixed during the evaluation (optional).
\end{description}
Each entry in the \yamlkey{observables} list is itself a map with a mandatory \yamlkey{name} key --- the qualified name of the observable,
which may carry its own options --- and an optional \yamlkey{kinematics} key.
As a shorthand, \yamlkey{kinematics} may be given a list of kinematic specifications, in which case one observable is created per specification.

The following example predicts the binned $q^2$ distribution of the decay $D^+\to \bar{K}^0 e^+\nu_e$, expressed through the probability \yamlvalue{D->Klnu::P(q2\_min,q2\_max)} for $q^2$ to fall within each of eight bins.

\begin{lstlisting}[language=YAML]
predictions:
  - name: PDF-Dp-to-K-e-nu
    observables:
      - name: D->Klnu::P(q2_min,q2_max);l=e,q=d,form-factors=SE
        kinematics:
          [{q2_min: 5.0e-6, q2_max: 0.2}, {q2_min: 0.2, q2_max: 0.4}, {q2_min: 0.4, q2_max: 0.6},
           {q2_min: 0.6,    q2_max: 0.8}, {q2_min: 0.8, q2_max: 1.0}, {q2_min: 1.0, q2_max: 1.2},
           {q2_min: 1.2,    q2_max: 1.4}, {q2_min: 1.4, q2_max: 1.6}]
\end{lstlisting}

\begin{figure}[t]
    \centering
    \includegraphics[width=0.6\textwidth]{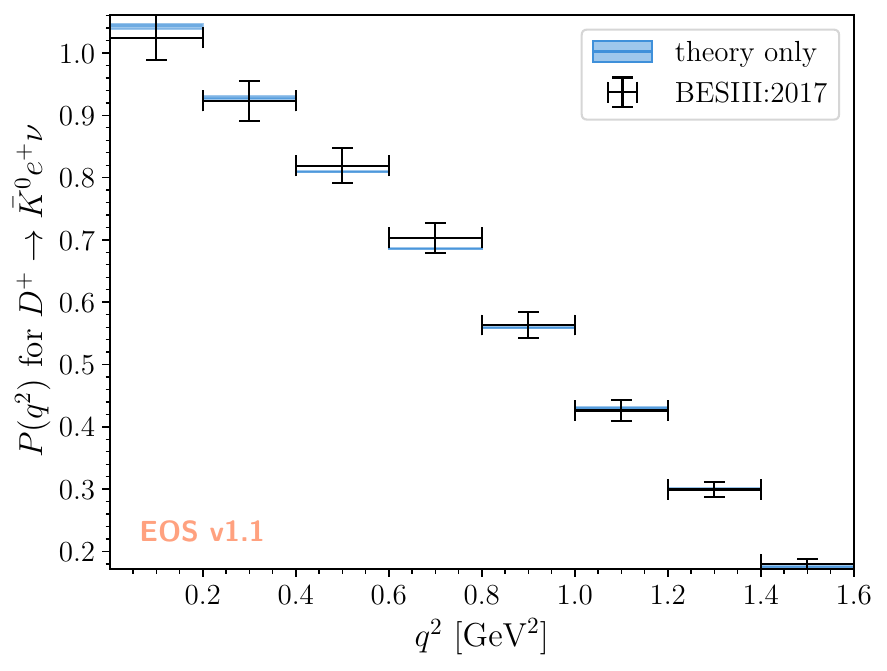}
    \caption{%
        Posterior-predicted binned $q^2$ distribution of $D^+\to \bar{K}^0 e^+\nu_e$,
        obtained from the \texttt{PDF-Dp-to-K-e-nu} prediction of the example analysis.%
    }
    \label{fig:pdf-dp-to-k-e-nu}
\end{figure}

The resulting binned distribution, including the uncertainties propagated from the posterior, is shown in \cref{fig:pdf-dp-to-k-e-nu}.

\subsection{Sample Masks}
\label{sec:analysis-files:masks}

A posterior need not be unimodal. Well-separated modes arise for a variety of reasons: an exact
symmetry of the likelihood, an approximate one, or a degeneracy of the available constraints that
carries no deeper meaning. Whatever their origin, a prediction that averages over all modes is not always meaningful.
The optional \yamlkey{masks} key therefore defines named \emph{masks}, each of which selects a subset of the
importance samples of a posterior. Each mask carries a \yamlkey{name}, a \yamlkey{description}, i.e., the
ordered list of its entries, and an optional \yamlkey{logical\_combination}, either \yamlvalue{and} (the default)
or \yamlvalue{or}. A sample is retained if all entries (for \yamlvalue{and}) or any entry (for \yamlvalue{or}) evaluate to a positive value.
An entry is either an existing observable, given by its qualified \yamlkey{name}; a new (pseudo-)observable
defined inline through an \yamlkey{expression}, using the syntax of \cref{sec:analysis-files:observables};
or another mask, referenced by \yamlkey{mask\_name}, so that masks can be composed.

Although not part of this analysis example, the analysis of nonleptonic $\bar{B}_{(s)}\to D_{(s)}^{(*)}\{\pi,K\}$
decays in Ref.~\cite{Meiser:2024zea} illustrates the use of masks.
Its \yamlvalue{WET-2} posterior, which varies the Wilson coefficients $\mathcal{C}_5$ through $\mathcal{C}_{10}$
of the $b\to c\bar{u}q$ sector, exhibits two well-separated modes A and B, whose local evidences differ by a
factor of about eight. They are separated by the two parallel hyperplanes
$f_{1,2} = \mathcal{C}_5 + 0.81\,\mathcal{C}_6 \mp 2.25$: mode A satisfies $f_1 < 0$ and $f_2 > 0$,
whereas mode B satisfies $f_1 > 0$. The original analysis separated the modes by writing custom scripts;
the following masks select them within the analysis file.

\begin{lstlisting}[language=YAML]
masks:
  - name: WET-2-mode-A
    description:
      - { name: 'mask::A1', expression: '2.25 - [[qbcu::Re{c5}]] - 0.81 * [[qbcu::Re{c6}]]' }
      - { name: 'mask::A2', expression: '2.25 + [[qbcu::Re{c5}]] + 0.81 * [[qbcu::Re{c6}]]' }

  - name: WET-2-mode-B
    description:
      - { name: 'mask::B',  expression: '[[qbcu::Re{c5}]] + 0.81 * [[qbcu::Re{c6}]] - 2.25' }
\end{lstlisting}

The \code{create-mask} task evaluates a mask on the samples of a posterior and stores the result as an
\pyclass{eos.data.SampleMask} object. The \code{predict-observables} task accepts the mask through its
\yamlkey{mask\_name} argument and propagates only the selected samples, which yields, e.g., the branching
ratios of the \yamlvalue{BRs} prediction restricted to a single mode.

\subsection{Custom Parameters}
\label{sec:analysis-files:parameters}

The optional \yamlkey{parameters} key defines custom parameters within the scope of the analysis.
Each entry is keyed by the qualified name of the new parameter and accepts the keys \yamlkey{central},
\yamlkey{min}, \yamlkey{max}, \yamlkey{unit}, and \yamlkey{latex}, i.e., the same information that describes any built-in \EOS parameter.
Besides introducing genuinely new parameters (e.g., for use in a custom observable; see \cref{sec:analysis-files:observables}),
this mechanism can also define a parameter as an \emph{alias} of one or more existing parameters through the optional \yamlkey{alias\_of} key.
Varying such an alias simultaneously varies all aliased parameters, providing a convenient means to impose symmetries among them.

In the example analysis, aliases are used to encode lepton-flavor universality (LFU) of the \WET Wilson coefficients.
\EOS assigns each lepton flavor $\ell\in\{e,\mu,\tau\}$ its own set of $c\to s\ell^+\nu_\ell$ Wilson coefficients, distinguished
by the parameter-name prefixes \code{scnuee}, \code{scnumumu}, and \code{scnutautau}.
Aliasing the corresponding coefficients onto a single flavor-universal parameter with the prefix \code{scnull} ties them together,
so that the fit varies them as one and thereby enforces LFU.
The following example does so for the real part of the left-handed vector coefficient $\mathcal{C}_{V_L}$.

\begin{lstlisting}[language=YAML]
parameters:
  'scnull::Re{cVL}':
      alias_of: [ 'scnuee::Re{cVL}', 'scnumumu::Re{cVL}', 'scnutautau::Re{cVL}' ]
      central:  1.0
      min:      0.88
      max:      1.03
      unit:     '1'
      latex:    '\mathrm{Re}\, \mathcal{C}^{\bar{s}c\bar{\nu}_\ell\ell}_{V_L}'
\end{lstlisting}

\subsection{Custom Observables}
\label{sec:analysis-files:observables}

The optional \yamlkey{observables} key defines custom observables within the scope of the analysis.
Each entry is keyed by the qualified name of the new observable and accepts the keys \yamlkey{latex} (its \LaTeX{} representation),
\yamlkey{unit}, and \yamlkey{expression}, as well as an optional \yamlkey{options} map.
The \yamlkey{expression} defines the new observable in terms of existing ones, whose qualified names are enclosed in double
angle brackets; parameters are referenced in double square brackets instead. An observable reference that carries no
options falls back to the parameter of the same name, so that \code{<<decay-constant::D_s>>} and
\code{[[decay-constant::D_s]]} are equivalent. We refer to the \EOS documentation~\cite[\href{https://eoshep.org/doc/releases/v1.1/reference/defining-observables.html}{Reference Manual/Defining Observables}]{EOS:doc} for the full expression syntax.
As noted above, loading the analysis file registers all custom observables that it defines, so that they can subsequently
be used like any built-in observable, for instance within a prediction or a manual constraint.

In the example analysis, one such observable is the ratio of the tensor $D_s^*$ decay constant to the $D_s$ decay constant,
$f_{D_s^{*,T}}/f_{D_s}$, which enters the \yamlvalue{TH-Dsstar-QCDSR} likelihood as a manual constraint.

\begin{lstlisting}[language=YAML]
observables:
  'decay-constant::D_s^*,T/D_s':
    latex: '\frac{f_{D_s^{*,T}}}{f_{D_s}}'
    unit: '1'
    options: {}
    expression:
      '<<decay-constant::D_s^*,T>> / <<decay-constant::D_s>>'
\end{lstlisting}

\subsection{Forward-Declared Constraints}
\label{sec:analysis-files:constraints}

The optional \yamlkey{constraints} key contains a list of forward-declared constraints.
Such a constraint does not yet exist when the analysis file is written; it is produced in the course of the analysis,
e.g., by the \code{create-constraint} task, which fits a multivariate Gaussian distribution to the samples of a posterior.
Each declaration has three mandatory keys:
\begin{description}
    \item[\yamlkey{name}] The qualified name under which the constraint is registered (mandatory).
    \item[\yamlkey{type}] The type of the declaration (mandatory). Presently, only \yamlvalue{external} is supported.
    \item[\yamlkey{filename}] The path to the file that holds the constraint's description, relative to the base directory of the analysis (mandatory).
\end{description}
The file must contain a key/value map from the constraint's name to its description, i.e., the same format as used by the built-in constraints database.
By convention, it resides in the \code{constraints/} subdirectory of the base directory, which is also where the \code{create-constraint} task
stores its output.
Before a posterior is constructed, \EOS registers each declared constraint whose file already exists.
A declared constraint can subsequently be referenced like any built-in constraint, e.g., in a prior or a likelihood component.

Although not part of this analysis example, the following listing illustrates how the $D\to K$ form-factor parameters inferred from
the \yamlvalue{CKM-D-to-K-l-nu} posterior can be reused as a correlated prior in a further posterior.
\begin{lstlisting}[language=YAML]
constraints:
  - name: 'D->K::FormFactors[SE]@CKM-D-to-K-l-nu'
    type: external
    filename: 'constraints/D->K::FormFactors[SE]@CKM-D-to-K-l-nu/constraint.yaml'

priors:
  - name: FF-D-to-K-posterior
    descriptions:
      - { type: 'constraint', constraint: 'D->K::FormFactors[SE]@CKM-D-to-K-l-nu' }
\end{lstlisting}

\subsection{Reproducible Analysis Steps}
\label{sec:analysis-files:steps}

The \yamlkey{steps} key contains a list of steps, each comprising predefined tasks, that reproduce an analysis. Each step contains three mandatory and two optional keys:

\begin{description}
    \item[\yamlkey{title}] The title of the step (mandatory).
    \item[\yamlkey{id}] The unique identifier of this step within the analysis (mandatory).
    \item[\yamlkey{tasks}] The list of tasks within the step (mandatory).
    \item[\yamlkey{depends\_on}] A list of step \yamlkey{id}s that must be executed before this step (optional).
    \item[\yamlkey{default\_arguments}] A key/value map from task names to default arguments for the tasks of this step (optional).
\end{description}

Each entry in the \yamlkey{tasks} list is a map with an \EOS-defined \yamlkey{task} and the set of \yamlkey{arguments} it requires. We refer to the \EOS documentation~\cite[\href{https://eoshep.org/doc/releases/v1.1/reference/python.html\#common-tasks}{Reference Manual/Python API}]{EOS:doc} for the arguments of each task.

The following example illustrates two nested-sampling steps, each followed by a step that draws the
corner figure of the resulting posterior samples. The figures themselves are declared under the
\yamlkey{figures} key and are referenced by name, so that a step only records which figure to draw
and in which formats.
Defining \yamlkey{steps} in this way ensures that any repeated sampling uses the same hyperparameters as those used to derive the initial results of the analysis.

\begin{lstlisting}[language=YAML]
figures:
  - name: CKM-all-corner
    type: 'corner'
    contents:
      - path: 'data/CKM-all/samples'
        label: 'CKM-all'

  - name: WET-all-corner
    type: 'corner'
    contents:
      - path: 'data/WET-all/samples'
        label: 'WET-all'

steps:
  - title: 'Sample from CKM-all posterior'
    id: 'CKM-all.sample'
    tasks:
      - task: 'sample-nested'
        arguments:
          posterior: 'CKM-all'
          bound: 'multi'
          nlive: 100
          dlogz: 9.0
          maxiter: 4000

  - title: 'Sample from WET-all posterior'
    id: 'WET-all.sample'
    tasks:
      - task: 'sample-nested'
        arguments:
          posterior: 'WET-all'
          bound: 'multi'
          nlive: 100
          dlogz: 9.0
          maxiter: 4000

  - title: 'Draw the corner figure for the CKM-all posterior'
    id: 'CKM-all.draw-figure'
    depends_on: ['CKM-all.sample']
    default_arguments:
      draw-figure:
        format: ['pdf', 'png']
    tasks:
      - task: 'draw-figure'
        arguments:
          figure_name: 'CKM-all-corner'

  - title: 'Draw the corner figure for the WET-all posterior'
    id: 'WET-all.draw-figure'
    depends_on: ['WET-all.sample']
    default_arguments:
      draw-figure:
        format: ['pdf', 'png']
    tasks:
      - task: 'draw-figure'
        arguments:
          figure_name: 'WET-all-corner'
\end{lstlisting}

\subsection{Metadata}
\label{sec:analysis-files:metadata}

The optional \yamlkey{metadata} key provides bibliographic information about the analysis. It is a key/value map with four optional keys:
\begin{description}
    \item[\yamlkey{title}] A human-readable title of the analysis (optional).
    \item[\yamlkey{id}] A unique identifier of the analysis (optional).
    \item[\yamlkey{description}] A human-readable description of the analysis (optional).
    \item[\yamlkey{authors}] The list of authors of the analysis (optional).
\end{description}
Each entry in the \yamlkey{authors} list is a map with a mandatory \yamlkey{name} key and optional \yamlkey{affiliation} and \yamlkey{email} keys.
The metadata is made available to the template rendered by the \code{report} task; see \cref{sec:analysis-files:reporting}.

The following example illustrates the metadata of an analysis.
\begin{lstlisting}[language=YAML]
metadata:
  title: 'Exclusive c -> s l nu transitions'
  id: 'EOS-DATA-2024-01'
  description: 'Determination of |V_cs| and of the WET Wilson coefficients'
  authors:
    - name: 'First Author'
      affiliation: 'Important University'
      email: 'f.author@example.com'
\end{lstlisting}

\subsection{Validation}
\label{sec:analysis-files:validation}

An analysis file is a large, highly cross-referenced document, and mistakes in it are easy to make.
\EOS therefore validates an analysis file in three phases, which the \code{validate} task runs in turn.
The structural phase checks the shape of the \YAML document alone: it rejects unknown or missing keys and
malformed entries. The semantic phase resolves the file against itself and against the \EOS databases: it
checks that every posterior refers to priors and likelihoods that the file defines, that the qualified
names of observables, constraints, and parameters exist, and that no component is left unused. The optional
deep phase finally constructs every posterior and every set of predicted observables, which catches errors
that surface only at run time.

Each problem is reported as a diagnostic that carries a severity, either an error or a warning, and
identifies the offending item in two complementary ways: semantically, through the path of \YAML keys that
leads to it, and textually, through the line number at which it occurs in the source file. For example,
\begin{lstlisting}[language=bash]
line 472: posteriors/TH-all/fixed_parameters/D->K::tp@SE: Parameter 'D->K::tp@SE' is fixed
but used by neither the likelihood nor any prediction of posterior 'TH-all'
\end{lstlisting}
names both the posterior and the key within it that is at fault, and the line at which to correct it.

\subsection{Reporting}
\label{sec:analysis-files:reporting}

Once the steps of an analysis have been completed, its results are distributed across a tree of data files. The
\code{report} task turns them into a human-readable document. It renders a user-supplied template,
written in the \texttt{Jinja2} templating language, to a Markdown file and, optionally, converts the
latter to PDF.

The template accesses the recorded results through a single object of class \pyclass{eos.reporting.AnalysisData},
which discovers the posteriors by scanning the data directory. Each posterior is exposed as an object of
class \pyclass{eos.reporting.PosteriorData}, which provides its importance samples, its nested-sampling
results, its modes and the associated goodness of fit, and its posterior predictions, as well as a helper
that draws the corner figure of its samples. All of these are loaded lazily, so that a report reads only
the files it actually refers to.

For examples, we refer to the \EOS user guide~\cite[\href{https://eoshep.org/doc/releases/v1.1/user-guide/}{User Guide}]{EOS:doc}
and the supplementary material for version 1.1~\cite{EOS-DATA-2026-04}.

%
%
%
%
\section{Improving Mode Finding and Posterior Sampling}
\label{sec:sampling}

In version 1.0, \EOS performed mode finding and posterior sampling directly in the parameter space specified by the user, without applying any parameter transformations. This functionality was implemented in the \pyclass{eos.Analysis} class, which could be used either directly from scripts and notebooks or indirectly through the \pymodule{eos.tasks} module and the command-line interface.
With version 1.0.4, \EOS switched to dynamic
nested sampling~\cite{Higson:2018} as its default sampling algorithm.
To this end, \EOS relies on the \dynesty software~\cite{Speagle:2020,dynesty:v3.0.0}.

While conceptually straightforward, operating directly in parameter space has two important drawbacks:

\begin{enumerate}
\item Informative priors are often represented by Gaussian \acp{PDF}, either exactly or approximately. Since a Gaussian distribution has infinite support, the corresponding parameter space is unbounded. For practical reasons, earlier versions of \EOS truncated the allowed range of such parameters, typically to the interval $\pm 3\sigma$ around the prior mean. While the effect of this truncation is negligible for a single parameter, we encountered analyses with $\dim \vecth = \mathcal{O}(100)$ parameters where the cumulative effect manifested as a low sampling efficiency.

\item Prior-based importance sampling~\cite{Lee1989} becomes substantially less efficient as the mismatch between prior and likelihood grows.
This inefficiency arises because the importance weights concentrate on a small fraction of prior draws as the mismatch grows, causing the effective sample size to collapse and the variance of the resulting estimator to increase accordingly.
Although this effect is usually modest for low-dimensional problems, it can substantially increase computational costs when the number of parameters is large. A notable example is \texttt{EOS-ANALYSIS-2022-05}~\cite{Leljak:2023gna,EOS-DATA-2023-01v2}, which was only marginally feasible with the available computational resources.
\end{enumerate}

To address these limitations, the \pyclass{eos.Analysis} class was rewritten in version 1.0.9 without changing its interface. Mode finding and prior sampling for a $D$-dimensional posterior are now performed in the space of cumulative prior probabilities (CPP space), denoted by
\[
\vec{u} = (u_1,\ldots,u_D), \qquad 0 \le u_n < 1.
\]

For a univariate prior $\vartheta_n \sim P_n$, sampling proceeds by drawing
$
u_n \sim \mathcal{U}_{[0,1)}
$
and transforming these samples to parameter space using inverse transform sampling,
$
\vartheta_n = \operatorname{CDF}_{P_n}^{-1}(u_n),
$
where
\begin{equation}
    \operatorname{CDF}_{P_n}(\vartheta_n)
        =
            \int_{-\infty}^{\vartheta_n} dx \, P_n(x)\, .
\end{equation}

For uncorrelated multivariate priors, this procedure is applied independently to each factor of the prior \PDF. Correlated priors require additional treatment. Correlated multivariate Gaussian priors\footnote{%
    Besides the Gaussian, \EOS version 1.1 supports one further correlated multivariate prior,
    named \texttt{transform}: a uniform distribution on a linearly transformed and shifted box.
    It is treated analogously, by inverting the linear map before the CPP transformation is applied
    to each component. The procedure can readily be generalized to further non-Gaussian priors
    when the need arises.
}
are handled through a Cholesky decomposition of their covariance matrix, $\Sigma = L L^T$ with $L$ lower triangular. Applying $L^{-1}$ to the mean-subtracted parameter vector transforms the original correlated distribution into a product of independent standard normal distributions. The CPP transformation is then applied independently to these components.

\begin{figure}
    \centering
    \includegraphics[width=.85\textwidth]{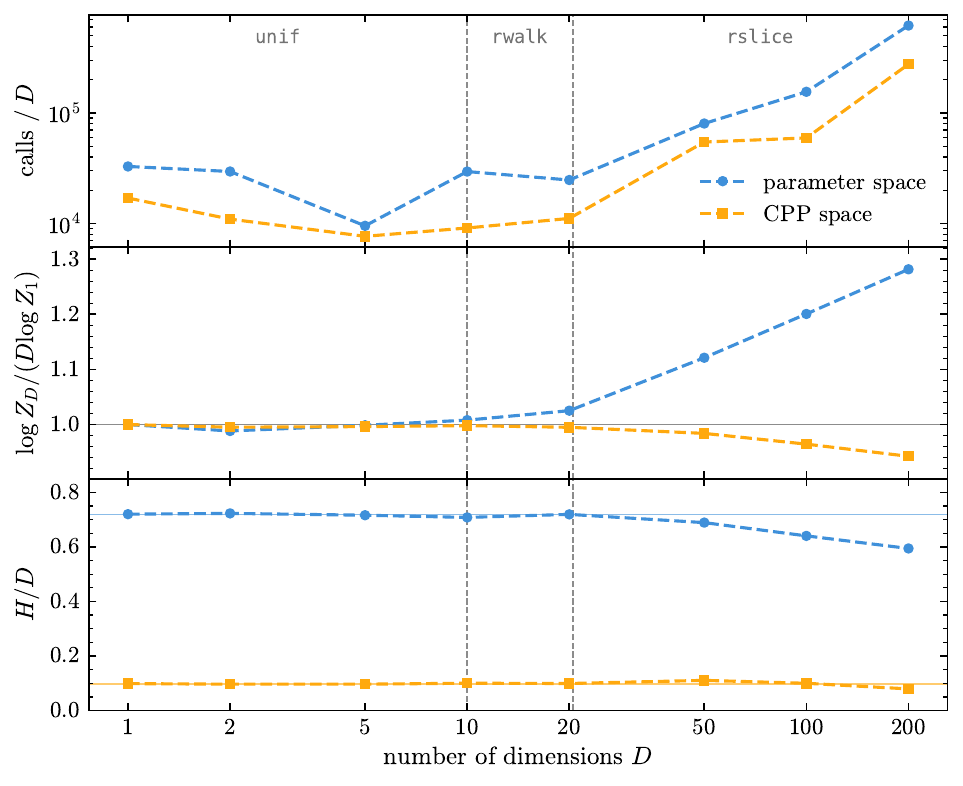}
    \caption{
        Comparison of the two sampling strategies as a function of the
        benchmark dimensionality. The upper panel shows the number of likelihood evaluations per
        dimension, the middle panel the relative normalized evidence $X(D)$, and the lower panel the
        information $H$ per dimension, i.e., the Kullback--Leibler divergence from the prior to the
        posterior; the thin horizontal lines in the lower panel give the exact values, $\simeq 0.7$ and
        $\simeq 0.1$ per dimension for the parameter-space and CPP-space implementations, respectively.
        Results obtained with the parameter-space implementation
        are shown as blue circles, those from the CPP-space implementation as orange squares.
        The dashed vertical lines mark the dimensionalities at which \dynesty changes its internal
        sampling method.
        \label{fig:sampling:comparison}
    }
\end{figure}

CPP space is bounded and corresponds to the hypercube $[0,1)^D$, where $D=\dim\vecth$. Operating in this space provides two important advantages.

First, numerical computations become more stable. The spacing between representable floating-point numbers grows with their magnitude, so parameters confined to $[0,1)$ are resolved more finely than those spread over large or unbounded domains, which improves numerical precision. Since \EOS does not provide analytical derivatives of the likelihood with respect to model parameters, optimization routines must rely on numerical approximations. Consequently, mode-finding algorithms benefit directly from the improved numerical stability of CPP space.

Second, posterior sampling becomes more robust and efficient. To quantify this improvement, we compare two
benchmarks, one sampling in parameter space as in earlier versions of \EOS and one sampling in CPP space,
each for dimensionalities $D \in \{1,2,5,10,20,50,100,200\}$.
Both benchmarks select the internal sampling method automatically, as \EOS does by default; \dynesty then samples
uniformly within the bound below $D=10$ (\texttt{unif}), by a random walk for $10\le D\le 20$
(\texttt{rwalk}), and by random slices above $D=20$ (\texttt{rslice}).
For each value of $D$, we considered a posterior constructed from a standard-normal likelihood and a standard-normal prior,
\begin{equation}
    \label{eq:sampling:model}
    P_D(\vecth)
        =
            \prod_{i=1}^{D}
            \underbrace{\mathcal{N}(\vartheta_i \mid \mu_i=0,\sigma_i=1)}_{\text{prior}}
            \,\cdot\,
            \underbrace{\mathcal{N}(\vartheta_i \mid \mu_i=0,\sigma_i=1)}_{\text{likelihood}},
\end{equation}
for which no parameter correlations are present, and in which prior and likelihood are perfectly aligned. We return to the
consequences of a mismatch between the two below. The accuracy of the evidence is benchmarked through the relative normalized evidence
\begin{equation}
    X(D)
        =
            \frac{\log Z_D}{D \, \log Z_{1}}
    \qquad
    \text{with }
    Z_D \equiv \int d^D\vartheta P_D(\vecth),
\end{equation}
which should satisfy $X(D)\simeq 1$ independently of $D$. Since the model factorizes over dimensions, $Z_D = Z_1^D$ holds
exactly for the model itself, irrespective of the truncation that earlier versions of \EOS applied; any deviation of $X(D)$
from unity therefore measures the error of the sampler alone.
To emulate the behavior of earlier \EOS versions, the benchmark labeled ``parameter space'' replaces the prior by a uniform
distribution on the interval $[-3,+3]$, corresponding to the $\pm 3\sigma$ truncation that those versions applied, and carries the
Gaussian of \cref{eq:sampling:model} in the likelihood instead. Both benchmarks therefore describe the same posterior,
but compress a different prior volume, which is what the lower panel of \cref{fig:sampling:comparison} quantifies.
Both benchmarks drive \dynesty directly, with the prior transform and the likelihood written out explicitly, rather
than through the \pyclass{eos.Analysis} class, so that the comparison isolates the effect of the parameter transformation
from the remainder of the \EOS machinery.

The benchmark results are shown in \cref{fig:sampling:comparison}. The upper panel shows the number of likelihood
evaluations per dimension. For both implementations this quantity is roughly flat up to $D \simeq 20$, with fluctuations that track the
changes of the internal sampling method, and grows steeply beyond, so that the total number of evaluations becomes
superlinear in the dimensionality only in the high-dimensional regime. Throughout the benchmark range, nested sampling
in CPP space consistently requires fewer
likelihood evaluations than the previous implementation, by a factor between approximately $1.2$ and $3.2$
depending on dimensionality.

The middle panel compares the estimated evidences. The CPP-space implementation produces values much closer to the expected result $X(D)=1$ across the full benchmark range. In particular, the logarithm of the evidence remains accurate to within $6\%$ up to $D = 200$, where the parameter-space result is already off by $28\%$.
To improve on this figure of merit, we are actively investigating how to modify the stopping criterion used by the \dynesty sampler to decide when to terminate sampling.

The lower panel shows the information $H = \mathrm{KL}(\text{posterior}\,\|\,\text{prior})$ per dimension,
the quantity that governs both the cost and
the accuracy of nested sampling: the posterior occupies a fraction $e^{-H}$ of the prior mass, so that
the sampler requires $\mathcal{O}(n_\mathrm{live} H)$ iterations and incurs an error on $\log Z$ that
scales as $\sqrt{H/n_\mathrm{live}}$~\cite{Ashton:2022grj}. Sampling in CPP space reduces $H$ from $0.72$ to $0.097$ per
dimension, since the prior of the CPP-space implementation is far closer to the posterior than the
truncated interval of the parameter-space implementation. Over $D = 100$ dimensions this is a
difference of some $27$ orders of magnitude in the prior volume that has to be compressed, and it is
the origin of the improvement seen in the upper two panels. The measured values agree with these
expectations up to $D = 20$; beyond that the parameter-space implementation reports a value of $H$
that falls increasingly short of the exact one, reaching $0.59$ instead of $0.72$ per dimension
at $D = 200$, whereas the CPP-space value departs only at the largest dimensionality shown. This is
a further symptom of the deterioration that the middle panel exhibits.

The benchmark above aligns prior and likelihood. Since a tension between the two is common in phenomenological
analyses, we repeat it with the likelihood shifted to $\mathcal{N}(\vartheta_i \mid \mu, \sigma_i = 1)$
while the prior stays centered at zero, so that both implementations continue to describe the same posterior.
\Cref{tab:sampling:misalignment} collects the results at $D = 100$, quoting the error of $\log Z$ per dimension
rather than $X(D)$: the latter is a ratio of logarithms and is ill-conditioned wherever $\log Z_1$ approaches zero,
which the parameter-space benchmark does in this range.

The two figures of merit respond differently. The advantage in the number of likelihood evaluations erodes as the
mismatch grows, and the accuracy of the CPP-space implementation degrades with it; its error exceeds the parameter-space
error between $\mu = 1$ and $\mu = 2$. This behavior is expected. The work performed by a
nested sampler is set by $H$. A flat prior is insensitive to where the likelihood sits, so $H$ stays
constant as $\mu$ grows. For the Gaussian prior underlying CPP space it instead grows as $H = \mu^2/8 + \text{const}$
per dimension, since a likelihood concentrated in the tail of the prior confines the posterior to an exponentially
small fraction of the prior mass. Equating the two predicts a crossing at $\mu \simeq 2.2$, somewhat later than the
measured one; the scaling of $H$ accounts for the trend but not for its onset. Holding the
accuracy fixed therefore requires $n_\mathrm{live}$ to grow in proportion to $H$, whereas \EOS presently defaults to a
value proportional to $D$ alone.

Where prior and likelihood are in substantial tension, the advantage in the number of evaluations
shrinks and the evidence estimate requires a correspondingly larger number of live points.
We emphasize, however, that the benchmark shifts every parameter alike, so that a mismatch of $\mu$ standard deviations
per parameter amounts to a global discrepancy of $\mu\sqrt{D/2}$ standard deviations between prior and likelihood,
or some $14\sigma$ at $\mu = 2$ and $D = 100$. A posterior arising from a tension of this size is pathological in its
own right: such a fit would be rejected, and its outcome distrusted, well before the accuracy of the sampler would become
the limiting consideration. The large-$\mu$ rows of \cref{tab:sampling:misalignment} therefore stress-test the
algorithm rather than describe a regime in which we expect \EOS to be applied. We nevertheless intend to tie the
default choice of $n_\mathrm{live}$ to the information reported by the sampler in a future release.

\begin{table}
    \centering
    \begin{tabular}{crrrr}
        \toprule
              & \multicolumn{2}{c}{evaluations\,$/D$} & \multicolumn{2}{c}{$\log Z$ error\,$/D$}\\
        \cmidrule(lr){2-3}\cmidrule(lr){4-5}
        $\mu$ & param.\ space & CPP space & param.\ space & CPP space\\
        \midrule
        $0$   & $155\,600$ & $59\,500$  & $+0.13$ & $+0.04$\\
        $1$   & $154\,000$ & $87\,100$  & $+0.14$ & $+0.04$\\
        $2$   & $148\,600$ & $108\,200$ & $+0.15$ & $-0.35$\\
        $3$   & $177\,000$ & $106\,500$ & $+0.15$ & $-1.27$\\
        \bottomrule
    \end{tabular}
    \caption{%
        Effect of a mismatch of $\mu$ standard deviations per parameter between prior and likelihood, at $D = 100$.
        Shown are the number of likelihood evaluations per dimension and the error of $\log Z$ per dimension against
        the exact result, for the parameter-space and the CPP-space implementations.%
    }
    \label{tab:sampling:misalignment}
\end{table}

These benchmarks demonstrate that sampling in CPP space is both more efficient and more accurate than the
parameter-space approach used in earlier versions of \EOS, provided that prior and likelihood are broadly consistent.
This improvement makes it possible to apply \EOS reliably to the high-dimensional problems increasingly encountered in
global analyses. 

%
%
%
%
\section{Creating Figures with the New \texttt{eos.figure} Framework}
\label{sec:plotting}

In version 1.0, \EOS supported creating figures with classes provided in the \texttt{eos.plot} module.
This code provided a wrapper around the \matplotlib \Python package~\cite{matplotlib}. It had been written expressly
to create figures with a single set of axes and offered only rudimentary support for creating figures from
a \YAML source file. Support for modifying a figure or its axes was added only after version 1.0.
\smallskip

For version 1.1, we wrote a new \EOS-specific wrapper around \matplotlib that logically separates the
description of the whole figure, of the individual plots within it, and of the items displayed within each plot.
The new code is part of the \texttt{eos.figure} module.
Unlike the old code, the new code enables the user to draw \EOS-specific plots and individual items onto an
existing \matplotlib axes object. The framework thus remains a thin layer on top of the existing \matplotlib
functionality. This logical separation also makes it possible to write unit tests for \texttt{eos.figure}.
\smallskip

The new framework produces publication-quality figures from a declarative \YAML description, without the user
writing any plotting code. A figure description can be stored as a standalone \YAML file and rendered
interactively, or it can be included among the figures of an analysis file (see \cref{sec:analysis-files})
and rendered by the \code{draw-figure} task. The three-level logical separation described above is reflected directly
in the description format:
\begin{description}
    \item[\texttt{figure}] The top-level object; it manages the underlying \matplotlib figure and arranges one or more plots.
    \item[\texttt{plot}] A single set of axes (a ``canvas'') onto which content is drawn, including its axis labels and ranges, grid, and legend.
    \item[\texttt{item}] A single piece of content drawn onto a plot, such as an observable as a function of a kinematic
    variable, a data point with error bars, or a kernel density estimate of a set of samples.
\end{description}
A figure description is deserialized by the \pyclass{eos.figure.FigureFactory} class, which selects the concrete figure
class from the \yamlkey{type} key; each nested plot and item is likewise selected by its own \yamlkey{type} key.
Every figure description recognizes two top-level keys, \yamlkey{name} (the file-unique name that also determines the
output file name; mandatory only when embedding a figure within an analysis file) and \yamlkey{type} (the figure type, defaulting to \yamlvalue{single}).
The remaining keys depend on the chosen type.
\smallskip

\begin{lstfloat}[b!]
\begin{lstlisting}[%
    language=YAML
]
type: inset
plot:
  xaxis: { label: '$q^2$', unit: '$\textnormal{GeV}^2$', range: [0.35, 3.2] }
  yaxis: { label: '$P (\tau \to K_S \pi^- \nu)$', range: [0.0005, 15], scale: log }
  aspect: 0.618
  legend: { position: 'lower left' }
  grid: { visible: false }
  items:
    - { label: 'Belle 2007', type: 'constraint', constraints: 'tau->K_Spinu::P(q2_min,q2_max)@Belle:2007A',
        observable: 'tau->K_Spinu::P(q2_min,q2_max)', variable: 'q2' }
    - { label: 'nominal fit', type: 'uncertainty', datafile: 'data/data/FF_2_2_4_4/pred-tau-P', variable: 'q2' }
watermark: { position: 'lower center' }
inset:
  position: [0.52, 0.6]
  size: [0.45, 0.35]
  plot:
    xaxis: { range: [0.4, 0.68], ticks: { visible: true } }
    yaxis: { range: [0.01, 0.5], ticks: { visible: true } }
    items:
      - { label: 'Belle 2007', type: 'constraint', constraints: 'tau->K_Spinu::P(q2_min,q2_max)@Belle:2007A',
          observable: 'tau->K_Spinu::P(q2_min,q2_max)', variable: 'q2' }
      - { label: 'nominal fit', type: 'uncertainty', datafile: 'data/data/FF_2_2_4_4/pred-tau-P', variable: 'q2' }
\end{lstlisting}
\caption{
    Example description of an inset figure in \YAML format. This example uses data produced by the phenomenological study in Ref.~\cite{Kirk:2025ajn}, available in the supplementary material~\cite{EOS-DATA-2025-05}.
    \label{lst:plotting:example}
}
\end{lstfloat}

At the figure level, the \yamlkey{type} key selects one of \yamlvalue{single} (a single plot, the default),
\yamlvalue{inset} (a main plot with a smaller inset plot in one corner), \yamlvalue{grid} (a rectangular grid of plots), or
\yamlvalue{corner} (a triangular arrangement of one- and two-dimensional marginal distributions).
At the plot level, the \yamlkey{type} key selects either \yamlvalue{2D} (a two-dimensional plot on a single set of axes,
the default) or \yamlvalue{empty} (a blank placeholder, useful to leave a cell of a grid figure empty). A \yamlvalue{2D} plot
accepts optional keys that configure the appearance of its axes (\yamlkey{xaxis}, \yamlkey{yaxis}), the \yamlkey{grid}, the
\yamlkey{legend}, and a \yamlkey{title}. At the item level, the \yamlkey{type} key selects among a large set of content types:
theory predictions as a function of a kinematic variable or a parameter (e.g., \yamlvalue{observable}, \yamlvalue{expression},
\yamlvalue{uncertainty}, \yamlvalue{signal-pdf}); visualizations of statistical samples read from a data file
(e.g., \yamlvalue{histogram1D}, \yamlvalue{kde1D}, \yamlvalue{histogram2D}, \yamlvalue{kde2D}, \yamlvalue{contours2D}); experimental and
theoretical constraints from the built-in \EOS database (e.g., \yamlvalue{constraint}, \yamlvalue{constraint2D}); and simple
annotations (e.g., \yamlvalue{point}, \yamlvalue{errorbars}, \yamlvalue{band}, \yamlvalue{vertical}). Besides their type-specific keys,
all items share a set of optional appearance keys: \yamlkey{label}, \yamlkey{color}, \yamlkey{alpha}, \yamlkey{linestyle}, and
\yamlkey{linewidth}. The classes that back every key, and that document the meaning of each key in detail, are listed
among the \code{eos.figure} classes in the \EOS documentation~\cite[\href{https://eoshep.org/doc/releases/v1.1/reference/python.html\#module-eos.figure}{Reference Manual/Python API/Plotting Classes}]{EOS:doc}.
\smallskip

As an example, the description in \cref{lst:plotting:example} creates an inset figure: a figure with a main set of axes (the \yamlkey{plot}
part) covering the full area, and a smaller inset set of axes (the \yamlkey{inset} part) that magnifies a region of
interest. Here, both sets of axes display the normalized $q^2$ distribution of the decay $\tau^-\to K_S\pi^-\nu_\tau$,
overlaying the nominal fit of Ref.~\cite{Kirk:2025ajn} on the \texttt{Belle:2007A} constraint from the \EOS database;
the inset magnifies the low-$q^2$ region. The \yamlkey{position} and \yamlkey{size} keys of the
inset give the bottom-left corner and the extent of the inset in normalized figure coordinates.
\smallskip

\begin{figure}
    \centering
    \includegraphics[width=0.65\textwidth]{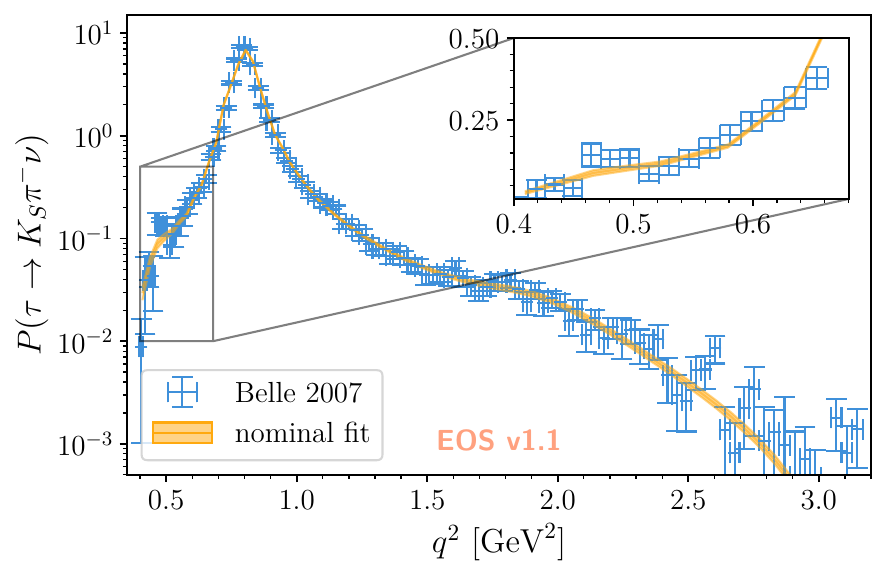}
    \caption{Example inset figure produced from the description in \cref{lst:plotting:example}.}
    \label{fig:plotting:example}
\end{figure}

A standalone figure description such as the one in \cref{lst:plotting:example} can be rendered in the command line using the
\texttt{eos-figure draw} command, or as part of an analysis file using the \texttt{eos-analysis draw-figure} command.
Within a custom \Python script or a \Jupyter notebook, it can be rendered through the \pyclass{eos.figure.FigureFactory} class.
Its \code{from\_yaml} method turns the \YAML description into a figure
object, whose \code{draw} method renders it. The separate \code{save} method then writes the rendered figure to one
or more output files, the format of each being determined by its file name extension. Drawing a figure a second time
draws its elements again on top of the first rendering; a figure should therefore be drawn only once, with all
output files passed to a single call to \code{save}.
\begin{lstlisting}[language=ipython]
import eos.figure

figure = eos.figure.FigureFactory.from_yaml(description)
figure.draw()
figure.save(['taudecay-inset.pdf', 'taudecay-inset.png'])
\end{lstlisting}
The result is shown in \cref{fig:plotting:example}.

%
%
%
%
\section{Summary and Outlook}\label{sec:summary}

We have presented version \EOSversion of the \EOS software and highlighted the changes
included since version 1.0.
In the five years that separate the two releases, the coverage of flavor observables has grown by
1063 (pseudo-)observables, which carry \EOS well beyond its original focus on semileptonic and rare
$b$-hadron decays. Nonleptonic $B$ decays, semileptonic and rare charm decays, hadronic $\tau$
decays, semileptonic kaon decays, and $e^+e^-$ scattering in the charmonium region are now available,
together with the local and nonlocal hadronic matrix elements that underpin them; we summarize these
additions in \cref{sec:new-observables}. This growth has been driven by the analyses it enables
rather than pursued for its own sake, and the software has in turn supported 21 peer-reviewed
publications as well as the interpretation of measurements by the Belle~II, CMS, and LHCb
collaborations.
\smallskip

The infrastructure has developed alongside the physics. Analyses are now described in a single,
human-readable analysis file, which records the priors, likelihoods, posteriors, predictions, and
figures of a study together with the steps that reproduce it, and which \EOS validates before it is
run; external likelihoods in the HistFactory format are accepted through this route as well. Mode
finding and posterior sampling operate in the space of the cumulative prior probability, which
removes the truncation that earlier versions applied to unbounded priors and, in our benchmarks,
reduces both the number of likelihood evaluations and the error on the evidence in the
high-dimensional problems that global analyses increasingly require. Figures are produced from a
declarative description through the new \texttt{eos.figure} framework, and the results of an analysis
can be collected into a report without writing bespoke code. As of this version, \EOS is
supported on macOS as well as on Linux.
\smallskip

The present release also points to the work ahead. The number of live points used in nested sampling
is tied to the dimensionality of the parameter space alone, although the cost and the accuracy of the
sampler are governed by the information gained from prior to posterior. We intend to tie the default
to the information that the sampler itself reports, and to revisit the condition under which nested
sampling terminates.
\smallskip

Finally, we commit to a stable interface for the releases that follow version 1.1, so that analyses
built on \EOS remain usable as the software continues to develop. Where the names and definitions of
parameters, constraints, and observables must change with our understanding of the physics, we record
those changes, as we do for the present release in \cref{app:breaking-changes}.

%
%
\section*{Acknowledgments}

We thank all \EOS contributors since v1.0 who are not authors of this article, including Frederik Beaujean, Christoph Bobeth, Maximilian Hoverath, and Jonas Eschle. DvD \& MR are grateful to the Mainz Institute for Theoretical
Physics (MITP) of the Cluster of Excellence PRISMA+ (Project ID 390831469) for enabling us to complete a significant portion of this work.

The work of NCB, CB, NG, SMu, and DS is supported by the Deutsche Forschungsgemeinschaft (DFG, German Research Foundation) under Germany’s Excellence Strategy – EXC 3107 – Project-ID 533766364.
The work of NCB is supported by the Konrad-Adenauer-Stiftung.
The work of CB was supported by the DFG within the Emmy Noether Program under grant number MI 2869/1-1.
The work of LG is supported by the Bundesministerium für Forschung, Technologie und Raumfahrt (BMFTR, German Federal Ministry of Research, Technology and Space) through the ErUM-Data project DEMOS (grant no. 05D25WM3).
NG has been funded by the DFG -- Emmy-Noether Grant No. 558599025.
FH is supported in part by the Swiss National Science Foundation
(SNF) under contract 200021-212729. 
The work of PL was supported by the DFG under grant 396021762 – TRR 257 “Particle Physics Phenomenology after the Higgs Discovery”.
The work of SMe is supported by the Studienstiftung des deutschen Volkes and the Spanish MCIN/AEI/10.13039/501100011033 through the following grants: grant CNS2022-135262 funded by the ``European Union NextGenerationEU/PRTR'', grant CEX2024-001451-M through the ``Unit of Excellence María de Maeztu'' award to the Institute of Cosmos Sciences, grant PID2022-136224NB-C21, and grant PID2025-169343NB-C22.
The work of SMu is supported by the Studienstiftung des deutschen Volkes and by the MKW NRW under funding code NW21-024-A.
The work of MBM and KKV is supported (in part) by the research programme Vidi with file number VI.Vidi.223.083, financed by the Dutch Research Council (NWO). In addition, KKV acknowledges support from the Dutch Research Council
(NWO) in the form of an Open Competition Domain Science grant. 
The work of DvD is supported by the UK Science and Technology Facilities Council (STFC) under
grants ST/V003941/1, ST/T001011/1, and ST/X003167/1.

%
%
%
%
%
\appendix
\makeatletter
\renewcommand\thesection{\@Alph\c@section}
\def\sectcounterend{}
\makeatother
\crefalias{section}{appendix}

\section{Breaking Changes since Version 1.0}
\label{app:breaking-changes}

We list here the changes since version 1.0 that may require an existing analysis to be adapted.
The list covers the entire range from version 1.0 to version \EOSversion. For the releases
from version 1.0.18 onward, it follows the \EOS changelog, which is part of the source code
repository~\cite{EOS:repo}; for the earlier patch releases of the 1.0 series, which predate the
changelog, it is based on a comparison of the names that \EOS registers in the two releases.
Purely additive changes, bug fixes, and changes internal to the implementation are not listed.

\subsection{Parameters}

\begin{itemize}
    \item The mass parameters of the hadronic matrix elements that are shared by the \texttt{BSZ2015},
    \texttt{SE}, \texttt{SSE}, \texttt{BGL1997}, and \texttt{HKVT2025} parametrizations have been renamed
    from an \code{@BSZ2015}/\code{@SE} suffix to \code{@HME}, since they are shared inputs rather than
    parametrization-specific fit results. Analysis files and stored parameter samples that name such a
    parameter explicitly have to be updated.

    \item The \LaTeX{} representation of a parameter must no longer contain \code{\$} signs.

    \item The Wilson coefficients of the $b\to s\nu\bar\nu$ sector have been renamed from
    \lstinline!sbnunu::Re{cL}! and \lstinline!sbnunu::Re{cR}! to \lstinline!sbnunu::Re{cVL}! and
    \lstinline!sbnunu::Re{cVR}!, and likewise for their imaginary parts, so that they match the
    naming of the scalar and tensor coefficients added alongside them.

    \item The parameters of three parametrizations have been removed together with the
    parametrizations themselves: the 43 nonlocal coefficients \code{@GRvDV2021}, the 10
    form-factor uncertainty parameters \code{@BZ2004} and \code{@BZ2004v2}, and the 7
    $B_s\to K^*$ coefficients \code{@FMvD2015}.
\end{itemize}

\subsection{Form-Factor Parametrizations and Models}

\begin{itemize}
    \item The parametrizations \texttt{ABR2022}, \texttt{BFW2010}, and \texttt{BMRvD2022} have been renamed
    to \texttt{SE}; see \cref{sec:changes}.

    \item The \code{form-factors} option of the semileptonic, rare, and nonlocal form-factor observables
    now defaults to \texttt{SSE}, instead of the previously process-dependent choices \texttt{BSZ2015},
    \texttt{BCL2008}, \texttt{BFvD2014}, \texttt{DM2016}, \texttt{SE}, and \texttt{DKMR2017}. Predictions
    made with the default form factors change accordingly; an analysis that relies on a previous default
    has to name it explicitly.

    \item The \texttt{BGL1997} parametrization of $P\to P$ form factors now imposes the exact relation
    $f_+(0) = f_0(0)$.

    \item The \code{WET-SMEFT} model has been removed. \EOS uses the most general basis of dimension-six
    \WET operators throughout; imposing the relations that follow from matching the Standard Model Effective Field Theory (SMEFT) onto the \WET
    is left to the user.
\end{itemize}

\subsection{Constraints}

The following constraints have been renamed:
\begin{itemize}\setlength{\itemsep}{0pt}
    \item \code{B->pi::FormFactors[parametric,LCSR+LQCD]@LvDM:2021A} $\to$ \code{...@LMvD:2021A}
    \item \code{B_s->D_s^*::A_1[s_max]@HPQCD:2019A} $\to$ \code{...@HPQCD:2019B}
    \item \code{B->K^*::A_12@FNAL+MILC:2013A}, \code{B->K^*::A_1@FNAL+MILC:2013A},
          \code{B->K^*::V@FNAL+MILC:2013A} $\to$ \code{...@HPQCD:2013B}
    \item \code{B_s->K^*::A_12@FNAL+MILC:2013A}, \code{B_s->K^*::A_1@FNAL+MILC:2013A},
          \code{B_s->K^*::V@FNAL+MILC:2013A} $\to$ \code{...@HPQCD:2013B}
    \item \code{D^+->Kbar^0e^+nu::BR[0.0,q2max]@PDG:2023A} $\to$ \code{...@PDG:2024A}
    \item \lstinline!0->pipi::Abs{f_+}^2@CMD3:2023C! $\to$ \code{...@CMD3:2023A}
    \item \code{B->K^*::FormFactors[A_0,A_1,A_2,V,T_1,T_2,T_23]@GRvDV:2021A} $\to$ \code{...@GRvDV:2022A}
    \item \code{B_(s)->D_(s)^*::FormFactors@HPQCD:2023A} $\to$ \code{...FormFactors[HQETBasis]@HPQCD:2023A}
    \item \code{B_(s)->D_(s)^*::FormFactors[TraditionalBasis]@HPQCD:2023A} $\to$ \code{...FormFactors@HPQCD:2023A}
\end{itemize}
Note that the last two entries reassign the name \code{B_(s)->D_(s)^*::FormFactors@HPQCD:2023A}, which now
refers to the traditional basis rather than the HQET basis; an analysis that uses either name has to
be checked rather than merely renamed.

In addition, the 10 constraints on the angular observables of $\bar{B}\to \bar{K}^*\psi$ decays express the
LHCb convention as an option in lieu of a name suffix, following the corresponding change to the
observables, e.g., \code{B^0->K^*0Jpsi::S_1s_LHCb@LHCb:2013E} $\to$ \code{B^0->K^*0Jpsi::S_1s@LHCb:2013E}.
The constraints \code{B^+->K^+l^+l^-::R_K[1.00,6.00]@LHCb:2021A}, \code{B^+->tau^+nu::BR@Belle:2015A},
\code{B^0_d->mu^+mu^-::BR@CMS+LHCb:2014A}, and the two
\code{B^0->(Dpi)^+l^-nu::KinematicalDistribution[...]@Belle:2017A} have been removed.

\subsection{Observables and Signal PDFs}

\begin{itemize}
    \item The signal PDFs have been renamed to a uniform \code{P(...)} scheme, e.g.,
    \code{B->pilnu::dGamma/dq2} $\to$ \code{B->pilnu::P(q2)},
    \code{B->D^*lnu::d^4Gamma} $\to$ \code{B->D^*lnu::P(q2,cos(theta_l),cos(theta_D),phi)}, and
    \code{B->Kll::d^2Gamma@LargeRecoil} $\to$ \code{B->Kll::P(q2,cos(theta_l))}. The changelog lists all
    17 renames.

    \item The signal PDFs \code{Lambda_b->Lambda_clnu::dBR} and \code{Lambda_b->Lambda_c(2625)lnu::dGamma}
    have been removed in favor of the corresponding multivariate PDFs. The unused PDFs
    \code{B->pimu1nu::d^2Gamma}, \code{B->pimu3nu::d^5Gamma}, \code{B->Dmu1nu::d^2Gamma},
    \code{B->Dmu3nu::d^5Gamma}, \code{B->Kll::d^2Gamma@LowRecoil}, and
    \code{B->K^*ll::d^2Gamma@LowRecoil} have been removed.

    \item The dedicated implementation of the $\bar{B}\to D^*\ell^-\bar\nu$ one-dimensional PDFs has been
    replaced by the general one, which renames \code{B->Dpilnu::P(...)} to \code{B->D^*lnu::P(...)} and
    updates the \texttt{Belle:2017A} constraints accordingly. The angular
    \code{B->D^*lnu::P(cos(theta_D)...)}, \code{...P(cos(theta_l)...)}, and \code{...P(phi...)} observables
    now return bin-averaged densities, consistent with the other one-dimensional PDFs.

    \item The polarization-dependent saturation observables \code{Saturation[...,0]},
    \code{Saturation[...,perp]}, and \code{Saturation[...,para]} of the $\Lambda_c\to\Lambda$ and
    $\Lambda_c\to n$ transitions have been removed; the polarization-summed \code{Saturation[...]} is to be
    used instead, as is already the case for the mesonic and $\Lambda_b$ transitions.

    \item The angular observables of $\bar{B}\to \bar{K}^{(*)}\ell^+\ell^-$, $\bar{B}_s\to\phi\ell^+\ell^-$, and
    $\bar{B}\to \bar{K}^*\psi$ decays express the LHCb convention as an option in lieu of a name suffix, e.g.,
    \code{B->K^*ll::S_1c_LHCb(q2)} $\to$ \code{B->K^*ll::S_1c(q2)@LHCb} and \code{B->K^*ll::A_FB_LHCb(q2)}
    $\to$ \code{B->K^*ll::A_FB(q2)@LHCb}; 32 observables are affected.

    \item The 11 explicitly CP-averaged observables carrying an \code{avg} suffix, such as
    \code{B->K^*ll::BRavg}, \code{B->K^*ll::A_FBavg}, and \code{B->Kll::F_Havg}, have been removed. The
    observables \code{BR}, \code{A_FB}, and \code{F_H} now denote the CP-averaged quantities themselves,
    which changes their meaning rather than only their name.

    \item The 51 accessors of the nonlocal form factors have been renamed to a uniform scheme, in which
    the modulus, the real part, and the imaginary part are expressed as \lstinline!Abs{...}!,
    \lstinline!Re{...}!, and \lstinline!Im{...}!, e.g., \code{B->K::abs_H_plus(q2)} $\to$
    \lstinline!B->K::Abs{H_plus}(q2)! and
    \code{B->K^*::re_ratio_long(q2)} $\to$ \lstinline!B->K^*::Re{ratio_long}(q2)!. The same applies to the
    \code{Hhat} and \code{ratio} accessors of $\bar{B}\to \bar{K}^{(*)}$ and $\bar{B}_s\to\phi$.

    \item The bespoke implementation of $\bar{B}_s\to \bar{K}^*\ell^-\bar\nu$ has been replaced by the generic
    $\bar{B}\to V\ell^-\bar\nu$ code. Its 35 observables have been removed in favor of the generic ones,
    among them the angular set \code{Shat_1c} to \code{Shat_6s}, \code{A_T^2} to \code{A_T^re},
    \code{H_T^1} to \code{H_T^5}, \code{R_long}, \code{R_para}, \code{R_perp}, \code{F_T}, and
    \code{F_L(q2)}; \code{B_s->K^*lnu::dBR/ds} is now \code{B_s->K^*lnu::dBR/dq2}.

    \item The angular signal PDFs of $\bar{B}\to D\pi\ell^-\bar\nu$ name their kinematic variables in full,
    \code{B->Dpilnu::P(c_D)} $\to$ \code{B->Dpilnu::P(cos(theta_D))}, and likewise
    \code{c_l} $\to$ \code{cos(theta_l)} and \code{chi} $\to$ \code{phi}, together with their binned counterparts.

    \item The 8 prior observables \code{b->c::Prior[...]@BGL} and \code{b->c::Prior[...]@CLN} have been
    removed, as have the 4 ratios \code{B->D^*::a_1/a_0[S_1]@HQE}, \code{...[V_1]@HQE},
    \code{B->D^*::a_2/a_0[S_1]@HQE}, and \code{...[V_1]@HQE}; the ratios for the remaining form factors
    are unaffected.

    \item The residue accessors of $B\to\pi\pi$ carry the parametrization as a suffix, e.g.,
    \lstinline!B->pipi::Im{Res{F_long}}(q2,k2)! $\to$ \lstinline!B->pipi::Im{Res{F_long}}(q2,k2)@FvDV2018A!,
    and the normalized transverse form factors drop their \code{_Normalized} suffix, e.g.,
    \code{B->K^*::F_long_T(q2)_Normalized} $\to$ \code{B->K^*::F_long_T(q2)}.

    \item The default numerical results of \code{B->gamma::F_V} and \code{B->gamma::F_A} (\texttt{FLvD2022QCDF})
    have changed with the implementation of the soft twist-5 and twist-6 contributions: at a representative choice
    of inputs, $F_V(2.16\,\GeV)$ moves from $0.2744$ to $0.2928$ and $F_A(2.16\,\GeV)$ from
    $0.2477$ to $0.2610$. In addition, the \code{contributions} option has been
    reworked. The value \code{soft} now selects all three soft terms rather than only the leading-twist soft
    term, which is now selected by \code{soft-nlo}. The value \code{partial-soft-tw-3+4} has been replaced by
    \code{soft-tw-3+4}.
\end{itemize}

\subsection{Kinematic Variables}

The kinematic variable of several decays has been renamed from \code{s} to \code{q2}, for consistency across
the software. This affects $\bar{B}\to P\ell^-\bar\nu$, $\Lambda_b\to\Lambda_c^{(*)}\ell^-\bar\nu$,
$D\to P\ell^+\nu$, $\Lambda_c\to (S=1/2)^+\ell^+\nu$, as well as $\bar{B}\to K^{(*)}\ell^+\ell^-$,
$\bar{B}_s\to\phi\ell^+\ell^-$, $\Lambda_b\to\Lambda^{(*)}\ell^+\ell^-$, and $\bar{B}\to X_s\ell^+\ell^-$.
Earlier in the 1.0 series the same rename was applied to
\code{Lambda_b->Lambda_c(2595)lnu::dBR/dsdtheta_l} and its $\Lambda_c(2625)$ counterpart.

\subsection{\Python Interface and Command-Line Clients}

\begin{itemize}
    \item The \code{eos.plot} package has been removed; use \code{eos.figure} instead. Along with it,
    the \code{corner-plot} task and the \texttt{eos-analysis corner-plot} and \texttt{eos-analysis plot-samples}
    commands have been removed, as have the \code{eos-plot}, \code{eos-plot-1d}, and \code{eos-plot-2d}
    scripts. A figure is now declared in the analysis file and drawn with the \code{draw-figure} task or the
    \texttt{eos-figure draw} command.

    \item The classes \pyclass{eos.Mutable} and \pyclass{eos.ParameterDescription} have been removed; no exported
    function or member could produce them.

    \item The scripts \texttt{eos-make-constraint}, \texttt{eos-merge-mcmc}, \texttt{eos-prepare-home},
    \texttt{eos-print-modes}, and \texttt{eos-print-uncertainty} have been removed, none of which provided
    useful functionality any longer; use \texttt{eos-analysis create-constraint} instead of
    \texttt{eos-make-constraint}. The
    \code{--base-directory} option of \texttt{eos-figure draw}, which the command never read, has been removed.

    \item The \cpp command-line clients \code{eos-evaluate}, \code{eos-print-polynomial}, and \code{eos-scan}
    have been removed together with the \code{libcli} library; polynomials in the Wilson coefficients are now
    created from \Python.

    \item Report templates now access the recorded results of an analysis through a single, lazily loaded
    \pyclass{eos.reporting.AnalysisData} object together with a \code{corner_figure} helper, instead of the
    previous \code{analyses}, \code{modes}, \code{len}, and \code{zip} template variables.

    \item The \pypmc package is now an optional dependency.
\end{itemize}

\subsection{Figures}

\begin{itemize}
    \item \pyclass{eos.figure.ConstraintResidueItem} now defaults to \code{style} $=$ \code{'pull'}, drawing each residue
    as a bar in units of standard deviations; the previous error-bar drawing of the absolute deviation
    remains available as \code{style} $=$ \code{'delta'}.

    \item An \pyclass{eos.figure.GridFigure} now stamps the \EOS watermark on a single plot (the bottom-rightmost
    non-empty plot) instead of every plot. Setting \code{watermark_plot} $=$ \code{'all'}
    restores the previous behavior.
\end{itemize}

\subsection{Data Files}

\begin{itemize}
    \item The varied parameters of an \pyclass{eos.data.MixtureDensity} are recorded as full parameter
    descriptions (name, minimum, maximum) under a \code{parameters} key, as the other \code{eos.data}
    objects already do, and its on-disk format has been bumped to version 2. Files in the previous format
    remain readable.

    \item The on-disk format of \pyclass{eos.data.Mode} has been bumped in the course of migrating the
    \code{eos.data} classes to the \pyclass{eos.Serializable}/\pyclass{eos.Deserializable} interface.
\end{itemize}

%
%
\bibliography{%
bibliography/cs.bib,%
bibliography/physics.bib,%
bibliography/statistics.bib,%
bibliography/lhcb.bib%
}
\bibliographystyle{bibliography/spphys}

\end{document}